\documentclass[twocolumn]{aastex6}
\usepackage{amsmath}
\usepackage{amssymb}
\usepackage{graphicx}
\usepackage[tight,hang]{subfigure}
\usepackage{natbib}
\usepackage{enumitem}


\newcommand{\msun}{\, \mbox{M}_{\odot}}
\newcommand{\muG}{\, \mu\mbox{G}}
\newcommand{\flash}{{\normalfont\scshape flash }}
\newcommand{\paramesh}{{\normalfont\scshape paramesh}}

\begin{document}

\title{The Evaporation and Survival of Cluster Galaxies' Coronae Part II: The Effectiveness of Anisotropic Thermal Conduction and Survival of Stripped Galactic Tails}
\author{Rukmani Vijayaraghavan\altaffilmark{1,2}}
\author{Craig Sarazin\altaffilmark{1}}
\email{rukmani@virginia.edu}
\altaffiltext{1}{Department of Astronomy, University of Virginia, 530 McCormick Rd., Charlottesville, VA 22904, USA}
\altaffiltext{2}{National Science Foundation Astronomy \& Astrophysics Postdoctoral Fellow}        

\begin{abstract}
We simulate anisotropic thermal conduction between the intracluster medium (ICM) and the hot coronal interstellar medium (ISM) gas in cluster galaxies. In the earlier Paper I \citep{Vijayaraghavan17b}, we simulated the evaporation of the hot ISM due to isotropic (possibly saturated) conduction between the ISM and ICM. We found that hot coronae evaporate on $\sim 10^2$ Myr timescales, significantly shorter than the $\sim 10^3$ Myr gas loss times due to ram pressure stripping. No tails of stripped gas are formed. This is in tension with the observed ubiquity and implied longevity of compact X-ray coronae and stripped ISM tails, and requires the suppression of evaporation, possibly due to magnetic fields and anisotropic conduction. We perform a series of wind tunnel simulations similar to Paper I, now including ISM and ICM magnetic fields. We simulate the effect of anisotropic conduction for a range of extreme magnetic field configurations: parallel and perpendicular to the ICM wind, and continuous and completely disjoint between the ISM and ICM. We find that when conduction is anisotropic, gas loss due to evaporation is severely reduced; the overall gas loss rates with and without anisotropic conduction do not differ by more than $10 - 20\%$. Magnetic fields also prevent stripped tails from evaporating in the ICM by shielding, and providing few pathways for heat transport between the ICM and ISM. The morphology of stripped tails and magnetic fields in the tails and wakes of galaxies are sensitive to the initial magnetic field configuration.
\end{abstract}

%
%

\maketitle

\section{Introduction}
\label{sec:intro}
The interstellar medium (ISM) gas of galaxies in dense galaxy clusters is subject to a range of violent astrophysical processes that remove, heat, and evaporate the gas. Cluster of galaxies are gravitationally bound regions of mass $10^{14} - 10^{15} \msun$, with most of the mass ($\gtrsim 85- 90\%$) in the form of collisionless dark matter. The dominant baryonic component ($\sim 10- 15\%$ of the mass) is the hot X-ray emitting intracluster medium (ICM) gas of temperature $T_{\rm ICM} = 10^7 - 10^8$ K and density $n_e = 10^{-2} - 10^{-4}$ cm$^{-3}$; galaxies and their stars comprise a small fraction ($\lesssim 2 - 5 \%$) of the total mass, as well as baryonic mass. The velocity dispersions of cluster galaxies are $\sim$ $10^3$ km s$^{-1}$. At these densities and velocities, ram pressure stripping can efficiently remove galactic interstellar medium (ISM) hot coronal gas and cold disk gas, a process that has been extensively studied using analytic calculations and numerical simulations \citep[e.g.,][]{Gunn72,Lea76,Gisler76,Larson80,Quilis00,Schulz01,Vollmer01,Roediger06,Kawata08,McCarthy08, Kapferer09,Tonnesen09, Ruszkowski14,Tonnesen14,Shin14,Roediger15a,Roediger15b,Vijayaraghavan15b}.

Thermal conduction between the ICM and ISM will lead to the heating of the ISM and its subsequent evaporation and dissipation into the ICM. In the earlier Paper I \citep{Vijayaraghavan17b} and this paper (Paper II), we describe the evolution of the hot ISM gas in an ICM-like environment with a focus on the effects of  isotropic and anisotropic thermal conduction. These processes are of particular interest in the context of the ICM and hot ISM gas, since the mean free path of ICM electrons $\lambda_e$ is comparable to the sizes of galaxies' hot gas coronae \citep{Spitzer62}:
\begin{equation}\label{eqn:lambdamfp}
\lambda_e = \frac{3^{3/2} (k_B T_e)^2}{4 \pi^{1/2} n_e e^{4} \ln \Lambda} \simeq 22~\textrm{kpc} \left(\frac{T_e}{10^8 \textrm{K}}\right)^2 \left(\frac{n_e}{10^{-3} \textrm{cm}^{-3}} \right)^{-1} \, ,
\end{equation}
where $T_e$ is the electron temperature, and the Coulomb logarithm $\ln \Lambda \approx 40$.
This can result in efficient heat transport to and subsequent dissipation of the hot ISM. This process is not necessarily isotropic since the hot ISM is threaded by $\muG$ magnetic fields  \citep[e.g.,][]{Vallee87,Clarke01,Carilli02,Govoni04,Kronberg05,Ryu12}, which force thermal conduction to be anisotropic.

Ram pressure stripping alone removes galaxies' hot ISM gas, over a range of galaxy orbits and masses, within an average timescale of $t_{\rm RPS} \lesssim 3$ Gyr \citep{Vijayaraghavan15b}.  We show in Paper I that fully saturated isotropic conduction results in the complete evaporation of galaxies' hot coronae, including the cooler, denser core, within $\sim$ $10^2$ Myr:
$t_{\rm ev} = 160$ Myr for a $2.8 \times 10^{11} \msun$ galaxy and $t_{\rm ev} = 500$ Myr for a $2.8 \times 10^{12} \msun$ galaxy. Thermal conduction and the evaporation of the ISM also result in a complete absence of stripped tails of ISM gas trailing the galaxy. The diffuse outer halo of the ISM gas, which when stripped forms these tails in numerical simulations that do not include conduction \citep[e.g.,][]{Toniazzo01,Acreman03,Shin14,Vijayaraghavan15b,Roediger15a,Roediger15b}, evaporates away before tails are even formed when conduction is allowed to effective. This result contradicts observations of stripped X-ray emitting gas tails
\citep[e.g.,][]{Forman79,Irwin96,Sun05b,Machacek06,Randall08,Kim08,Sun10,Kraft11,Zhang13} that are tens to hundreds of kpc long and have survived in the cluster environment for hundreds of Myr. 

We find in Paper I that after the outer diffuse coronae evaporate, central coronae themselves are heated and destroyed. X-ray observations contradict these results as well \citep[e.g.,][]{Vikhlinin01,Yamasaki02,Sun05c,Sun05,Sun07,Jeltema08}. These total evaporation timescales ($t_{\rm ev} \sim 10^2$ Myr) are significantly shorter than ram pressure stripping timescales ($t_{\rm RPS} \lesssim 3$ Gyr), and the observed abundance of stripped X-ray tails requires a mechanism balancing or suppressing evaporation due to thermal conduction. Radiative cooling can in principle precisely balance thermal conduction. Alternatively, magnetic fields will significantly reduce the effectiveness of thermal conduction: in the presence of a magnetic field, electrons are restricted from moving isotropically and gyrate around the direction of the local magnetic field. Electrons, and therefore the flow of heat through electrons, are restricted to move only along the local magnetic field, a process referred to as anisotropic thermal conduction. We discuss this process in more detail in \S~\ref{sec:methods}.

Previous studies have shown that anisotropic thermal conduction plays an important role in other ICM phenomena. Cold fronts observed at the interfaces between cores of merging subclusters and the surrounding ICM are likely in part shielded from the hotter ICM by local magnetic fields wrapped around the subclusters' leading edges, which prevent the diffusion of heat to cooler subcluster cores  \citep[e.g.,][]{Asai04,Asai05,Asai07,ZuHone13,Komarov14}. \citet{Asai04,Asai05} use two dimensional and three dimensional simulations to show that magnetic fields stretch and wrap around moving subclusters and suppress conduction, resulting in cold fronts that persist for $\sim 1$ Gyr. \citet{Asai07} find that this is true even for initially turbulent ICM magnetic fields. \citet{Lyutikov06} and \cite{Dursi08} show that magnetic fields perpendicular to the direction of motion of an infalling object like a subcluster or galaxy are amplified and wrapped around the object's leading edge; in this scenario, heat flow will be suppressed across magnetic field lines and therefore from the ICM to the ISM.  Anisotropic thermal conduction can also play a regulatory role in the temperature structure of the ICM. \citet{Binney81} suggested that thermal conduction suppressed by magnetic fields described the temperature structure of M87 -- cool gas surrounded by the hot ICM. In clusters with cool cores, the heat-flux-buoyancy instability \citep[HBI,][]{Balbus08,Quataert08,Parrish08,Parrish09,Bogdanovic09} can result in magnetic fields  being aligned in directions perpendicular to the temperature gradient, suppressing heat flow from the hot outer ICM to the cool core. However, \citet{Ruszkowski10,Ruszkowski11} showed that turbulent motions in the ICM can stir ICM magnetic fields resulting in the flow of heat from the outer ICM to the core. Thermal conduction can also smooth out density fluctuations in the ICM \citep{Gaspari13,Gaspari14}, with the level of smoothing sensitive to the extent to which magnetic fields can suppress conduction.

The strength of the ICM magnetic field has been observationally measured and constrained to be $B \sim \muG$ \citep[reviewed in ][]{Carilli02,Govoni04,Kronberg05,Ryu12}. Indirect evidence for the presence of cluster magnetic fields and their strengths comes from cluster radio halos produced by the synchrotron radiation of relativistic electrons gyrating around ICM magnetic fields \citep[e.g., ][]{Miley80,Giovannini93,Feretti99,Govoni04}. The integrated cluster magnetic field can be directly measured using the Faraday rotation measure (RM). As electromagnetic waves propagate through the ICM plasma, their plane of polarization is rotated, at a given wavelength $\lambda$, by $\Delta \chi = $~RM $\lambda^2$ \citep{Burn66}. The RM varies with the electron density, magnetic field strength, and pathlength $l$ as RM $= 0.81 \int n_e \mathbf{B} \cdot d\mathbf{l}$ rad m$^{-2}$. Faraday rotation measurements show that ICM magnetic fields are of $\sim \muG$ strengths coherent over scales of tens of kpc \citep[e.g., ][]{Vallee86,Vallee87,Kim90,Kim91,Taylor93,Taylor94,Clarke01,Taylor01,
Rudnick03,Murgia04,Bonafede10,Bonafede11,Govoni10,Vacca12,Bonafede13}. The morphology of ICM magnetic fields is more uncertain. A common assumption is that ICM magnetic fields are random and isotropic, with a Kolmogorov power spectrum, motivated by MHD turbulence power spectrum analyses \citep[e.g.,][]{Vogt03,Vogt05}. 
The magnetic field structure within the hot coronal component of the ISM is even more uncertain given the lack of star formation and cosmic rays, and therefore synchrotron emission not associated with AGN jets and radio lobes \citep{Beck11,Beck13}. The relative topology of the magnetic field for a galaxy's hot corona embedded in a cluster's ICM is therefore unknown. 

To account for these uncertainties in our series of \flash magnetohydrodynamic idealized numerical simulations exploring the effect of anisotropic thermal conduction between a realistic ICM and ISM, we consider extremes in orientation and relative structure of the ICM and ISM magnetic fields. We consider two variations in the direction of the magnetic field: perpendicular and parallel to the direction of the galaxy motion through the ICM; two variations in the relative topology of the ICM and ISM magnetic field: fully continuous and completely disjoint; and perform all of these simulations with and without anisotropic thermal conduction. Over this parameter space, we perform a total of $2 \times 2 \times 2 = 8$ simulations of a galaxy in a box with a magnetized ICM-like fluid. The mass of the galaxy ($M_{\rm galaxy}  = 2.8 \times 10^{11} \msun$) and the properties of the ICM ($\rho_{\rm ICM} = 2 \times 10^{-27}$ g cm$^{-3}$, $P_{\rm ICM} = 2 \times 10^{-11}$ dyne cm$^{-2}$, $\mathbf{v}_{\rm ICM} = 610~\hat{x}$ km s$^{-1}$) are identical to the fiducial model galaxy in Paper I.  The simulations without anisotropic thermal conduction are primarily used to calibrate our measurements of gas loss from evaporation due to thermal conduction.

This paper is structured as follows: We  describe our simulation methods and provide a brief theoretical description of anisotropic thermal conduction in \S~\ref{sec:methods}. The initial conditions of the simulated galaxy and the magnetic field orientation are described in \S~\ref{sec:ic}. We describe the results of the numerical simulations in \S~\ref{sec:results}. We then discuss the implications of these results for the survival of galactic coronae and stripped galaxy tails in \S~\ref{sec:discussion}. We summarize our conclusions in \S~\ref{sec:conclusions}.

\section{Simulation Methods} 
\label{sec:methods}
The numerical techniques used here are as in Paper I: we use the \flash 4.3 code \citep{Fryxell00, Dubey08}, a parallel, modular, grid-based magnetohydrodynamics code with adaptive mesh refinement (AMR, implemented using \paramesh; \citealt{MacNeice00}) to focus computational resources on the most interesting regions within a simulation. We use the multigrid solver \citep{Ricker08} to calculate the gravitational potential on the grid, and collisionless massive particles are mapped to the grid using cloud-in-cell interpolation. We use the unsplit staggered mesh (USM) solver \citep{Lee09, Lee13} to solve the equations of magnetohydrodynamics (MHD).  The USM solver uses a finite-volume, second-order Godunov method and a directionally unsplit scheme to solve the MHD governing equations. With the USM solver, we use the HLLC (Harten-Lax-van Leer-Contact; \citealt{Toro94, Li05}) Riemann solver  in \flash to calculate high order Godunov fluxes. \flash uses the constrained transport method of \citet{Evans88} to enforce the divergence-free constraint on magnetic fields, $\nabla \cdot \mathbf{B} = 0$. We use the \citet{Balsara01} prolongation method to ensure divergence-free prolongation of magnetic fields across mesh boundaries between cells at different refinement levels. 

In the absence of a magnetic field, thermal conduction is isotropic, and its  effectiveness depends on the local temperature gradient, temperature, and density. In the presence of a magnetic field, a charged particle like an electron experiences a Lorentz force, $\mathbf{F} = q_e (\mathbf{v_e} \times \mathbf{B)}$. For $\mathbf{v_e} \parallel \mathbf{B}$, the Lorentz force is zero. When $\mathbf{v_e} \perp \mathbf{B}$, the electron is subject to a Lorentz force that results in a circular motion around the magnetic field. The radius of this circle, known as the gyroradius or Larmor radius, is $r_g = m_e v_{e, \perp} c / q_e B $, where $v_{e, \perp}$  is the component of the electron's velocity perpendicular to the magnetic field. The typical gyroradius of an electron in our simulations is $\sim 10^7$ cm, while electron mean free paths are on the order of $\sim 1 - 10$ kpc. Therefore electrons (and ions) are effectively restricted to propagate only along directions parallel to the local magnetic field. Since thermal  conduction is carried out primarily by electrons in the ICM, in the presence of an ICM magnetic field, thermal conduction is no longer isotropic, and is effectively anisotropic.

The heat fluxes defined in Paper I are modified with the additional constraint that heat is only transported anisotropically. The modified Spitzer heat flux is:
\begin{equation}\label{eqn:conduction_aniso}
\mathbf{Q_{\rm Spitzer}} = -\kappa \hat{\mathbf{e}}_B (\hat{\mathbf{e}}_B \cdot \nabla T_e),
\end{equation}
where $\hat{\mathbf{e}}_B = \mathbf{B} / |\mathbf{B}|$, the unit vector along the local magnetic field direction.
The thermal conductivity is 
\begin{equation}
\kappa = \epsilon \delta_T \kappa_{\rm Spitzer},
\end{equation} where $\epsilon = 0.419$ and $\delta_T = 0.225$ are transport coefficients for a $Z = 1$ hydrogen plasma as derived in \citet{Spitzer53}, and 
\begin{equation}
\kappa_{\rm Spitzer} = 20 \left(\frac{2}{\pi}\right)^{3/2} \frac{k_B^{7/2} T_e^{5/2}}{m_e^{1/2} e^{4} Z \ln \Lambda}.
\end{equation}
$T_e$ is the electron temperature, $\ln \Lambda$ is the Coulomb logarithm, and $m_e$ and $q_e$ are the electron mass and charge.

We include the effects of saturation on the effective heat flux, following the \citet{Cowie77} prescription as in Paper I. Anisotropic conduction in \flash is implemented  with a monotonized central (MC) slope limiter based on the method described in \citet{Sharma07}, and used in \citet{Ruszkowski10} and \citet{ZuHone13}.

\subsection{Initial Conditions}
\label{sec:ic}
We simulate the evolution of a $2.8 \times 10^{11} \msun$ galaxy in an ICM-like wind tunnel with mass and hydrodynamic profiles identical to the galaxy in Paper I. The galaxy's total density profile a Navarro-Frenk-White profile \citep{Navarro97} with a virial mass $M_{\rm 200} = 2.8 \times 10^{11} \msun$. The gravitational potential is determined using collisionless dark matter particles whose positions and velocities correspond to the galaxy's profile. These particles have a mass of $10^7 \msun$ for a total of $3.3 \times 10^4$ particles (including the exponential fall off of the density profile at a radius $r > R_{200}$). Further details of the dark matter initialization are described in Paper I and in \citet{Vijayaraghavan17a}.

The mass of the galaxy's ISM is 10\% of the total galaxy mass and its density profile is a singular isothermal profile. The ICM has density $\rho_{\rm ICM} = 2 \times 10^{-27}$ g cm$^{-3}$, pressure $P_{\rm ICM} = 2 \times 10^{-11}$ erg cm$^{-3}$, and temperature $T_{\rm ICM} = 7.14 \times 10^7$ K. The incoming ICM wind has a velocity $\mathbf{v_{\rm ICM}} = v_{\rm ICM} \hat{x}$, with $v_{\rm ICM} = 610$ km s$^{-1}$. The galaxies are simulated in cubic boxes of side $2 \times 10^{24}$ cm, or $648$ kpc, with a maximum spatial resolution of 1.266 kpc corresponding to 6 levels of refinement.

\begin{figure*}[!htbp]
  \begin{center}
    \subfigure[$z = 0$ plane, $\mathbf{B} \parallel \mathbf{v_{\rm ICM}}$]
    {\includegraphics[width=0.49\textwidth]{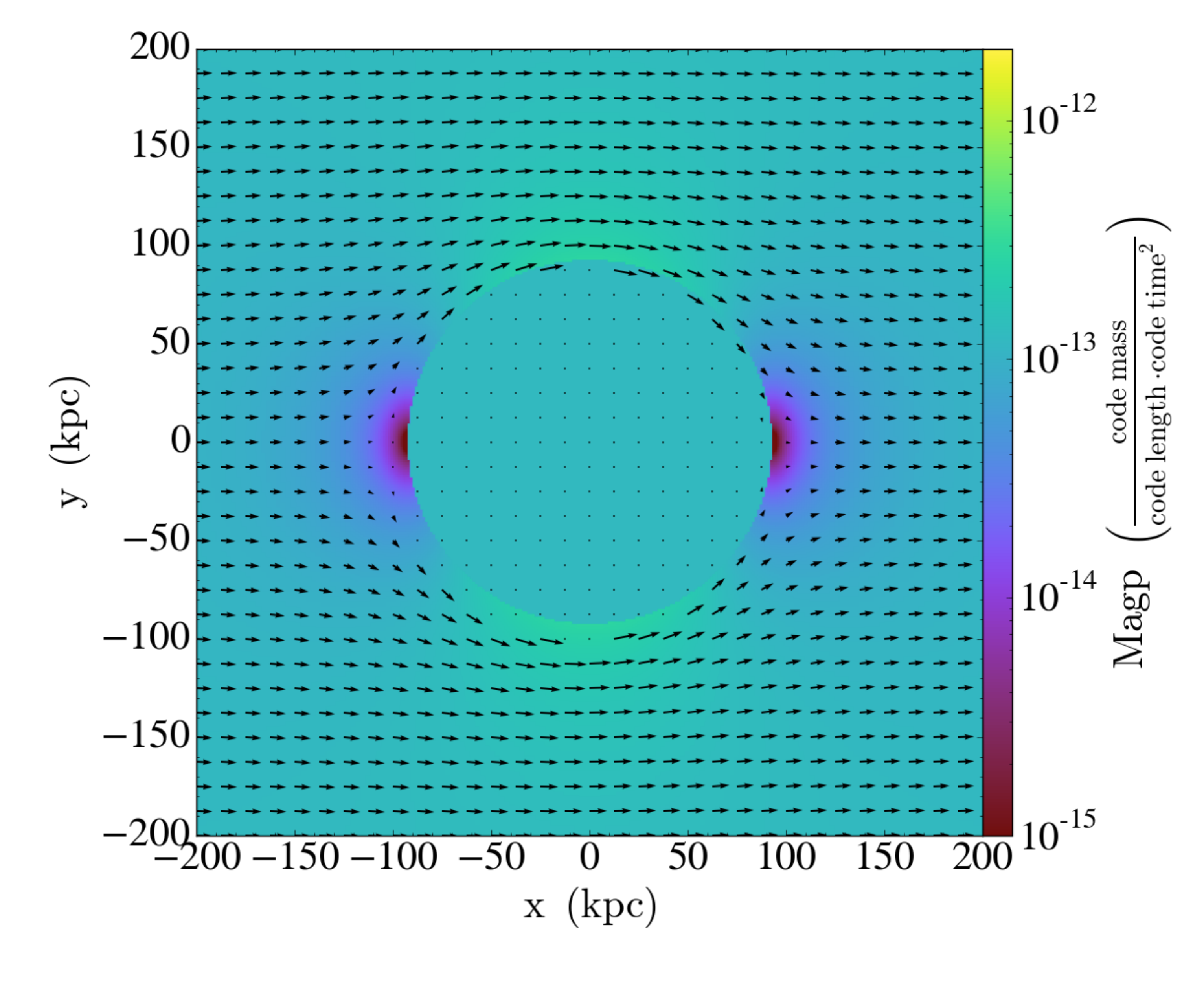}\label{fig:magpslc_z}}
    \subfigure[$y = 0$ plane, $\mathbf{B} \parallel \mathbf{v_{\rm ICM}}$]
    {\includegraphics[width=0.49\textwidth]{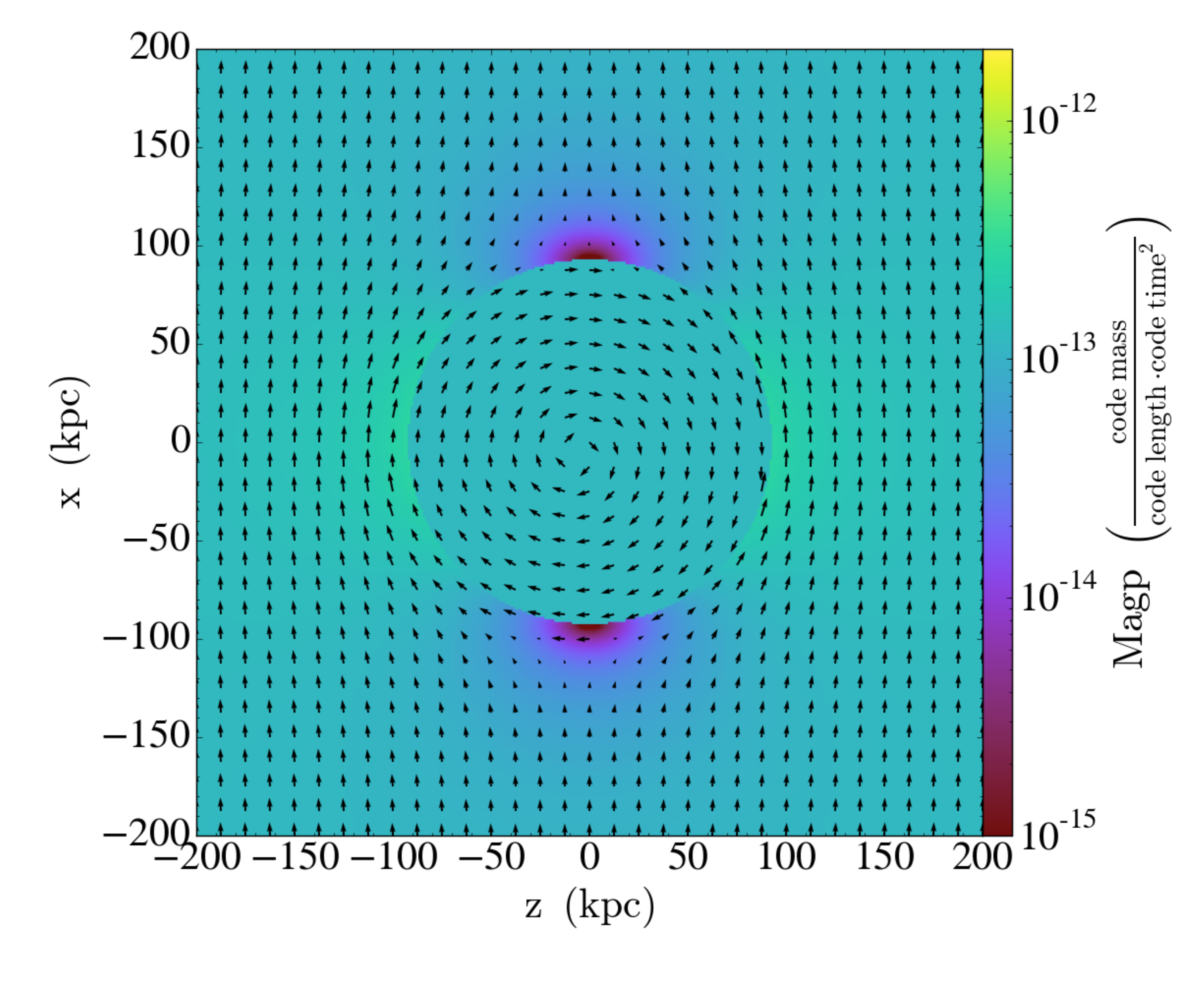}\label{fig:magpslc_y}}
     \caption{Slices in the $z = 0$ (left) and $y = 0$ (right) planes of the magnetic pressure with arrows denoting the magnetic field directions for the initial conditions when the ISM and ICM magnetic fields are disconnected.  The ICM field wraps around the galaxy, which shield the galaxy from heat conduction. The ISM magnetic field is toroidal in the $x-z$ plane, and therefore lies entirely in the plane of the figure in the $y = 0$ slice, and has no component in the $z = 0$ plane.
\label{fig:magpslc_ic}}
  \end{center}  
\end{figure*}

The ISM and ICM in these simulations are magnetized. We vary the morphology of the magnetic field in the various simulations. We perform two  classes of simulations: (1) with magnetic field lines continuous through the ICM and ISM, and (2) with a disjoint ICM-ISM magnetic field morphology, with the ICM magnetic field lines wrapped around the galaxy and a toroidal ISM magnetic field. In the first simulation series, we have a a uniform magnetic field in the simulation box with two possible orientations: $\mathbf{B} = B_0 \hat{x}$ and $\mathbf{B} = B_0 \hat{y}$, corresponding to directions parallel and perpendicular to the ICM wind. We adopt $B_{\rm FLASH} = 2 \sqrt{\pi} B_0 = 0.5 \muG$, corresponding to $B_0 = 1.77 \muG$, where $B_0$ is the magnetic field strength in CGS units, and $B_{\rm FLASH}$ is in internal \flash units. The initial magnetic pressure in the ICM is $P_{\rm B} = B^2/8\pi = 1.25 \times 10^{-13}$  erg cm$^{-3}$. The corresponding initial ICM plasma $\beta$ parameter is $\beta = P_{\rm thermal, ICM} / P_{\rm B, ICM} = 160$.

In the second simulation series, we modify the ICM magnetic field to wrap around the galaxy. Outside of the galaxy, the magnetic scalar potential in spherical polar coordinates is
\citep[][\S 5.12]{Jackson99}:
\begin{equation}
\Phi_M(r) = -B_0 r \cos \theta + \sum_{l = 0}^{\infty} \frac{\alpha_l}{r^{l+1}}P_l(\cos \theta),
\end{equation}
where $r$ is the galaxy-centric radius,
$B_0 = 1.77 \muG$ is the uniform magnetic field amplitude at $r \gg R_{200}$,
and the $P_l$'s are Legendre polynomials.
Here, $\theta$ is the polar angle measured from the direction of the field at large distances from the galaxy. When $\mathbf{B} \parallel \mathbf{v_{\rm ICM}}$, i.e., $\mathbf{B} = B_0 \hat{x}$ at $r \gg R_{200}$, and $\theta$ is the polar angle measured from the $x$-axis. When $\mathbf{B} \perp \mathbf{v_{\rm ICM}}$, i.e., $\mathbf{B} = B_0 \hat{y}$ at $r \gg R_{200}$, and $\theta$ is the polar angle measured from the $y$-axis. 
For the magnetic field outside the galaxy ($r > R_{200}$) to be disconnected with the field within the galaxy ($r < R_{200}$), the radial component of the field must go to zero at the boundary, $B_r = 0$ at $r = R_{200}$.
With this boundary condition,  the magnetic potential reduces to:
\begin{equation}
\Phi_M(r) = -B_0 \cos \theta \left(r + \frac{R_{200}^3}{2r^2}\right).
\end{equation}
The magnetic field, $\mathbf{B} = - \nabla \Phi_M$ is then:
\begin{equation}
\mathbf{B} = B_0 \cos \theta \left(1 - \frac{R_{200}^3}{r^3}\right) \hat{r} - B_0 \sin \theta \left(1 + \frac{R_{200}^3}{2 r^3}\right) \hat{\theta} \, ,
\end{equation}
where $\hat{r}$ and $\hat{\theta}$ are unit vectors in the direction of increasing $r$ and $\theta$, respectively.

We assume that the initial magnetic field within the galaxy is uniform in amplitude and toroidal:
\begin{equation}
\mathbf{B} = B_0 \hat{\phi} \, ,
\end{equation}
where $\phi$ is the azimuthal angle in $x-z$ plane
(measured from the $+x$ axis to the $+z$ axis)
in both the $\mathbf{B} \parallel \mathbf{v_{\rm ICM}}$ case and the $\mathbf{B} \perp \mathbf{v_{\rm ICM}}$ case. We performed low resolution test runs with varying toroidal field orientations -- i.e., with the internal magnetic field being $\mathbf{B} = B_0 \hat{\phi} $, but with $\phi$ as the azimuthal angle in the $y-z$ plane. This change in toroidal field orientation did not significantly affect our results and conclusions.

The configuration of the outer magnetic field is appropriately rotated to be consistent with the directions in the simulations in which the magnetic field is continuous between the ICM and ISM, i.e., so that $\mathbf{B} \parallel \mathbf{v_{\rm ICM}}$ or $\mathbf{B} \perp \mathbf{v_{\rm ICM}}$ as $r \gg R_{200}$. This disconnected ISM + ICM magnetic field configuration satisfies the solenoidal constraint, $\nabla \cdot \mathbf{B} = 0$, as does the uniform magnetic field.

The  shielded ICM field + toroidal ISM magnetic field configuration is illustrated in Figure~\ref{fig:magpslc_ic}, which shows zoomed-in slices of the magnetic pressure inside and in the immediate vicinity of the galaxy in the $z = 0$ and $y = 0$ plane when $\mathbf{B} \parallel  \mathbf{v_{\rm ICM}}$. Figure~\ref{fig:magpslc_z} shows the magnetic field in the $z = 0$ plane. Since the ISM magnetic field is toroidal in the $x - z$  plane, the field lines within the galaxy
are entirely perpendicular to the $x - z$ plane in this slice. The ICM magnetic field wraps around the galaxy, with ICM field lines squeezed and stronger magnetic pressure at $y = \pm R_{200}, x = 0$, and weak ICM magnetic pressure at diverging field lines near  $x = \pm R_{200}$, y = 0. Figure~\ref{fig:magpslc_y} shows field lines in the $y = 0$ plane; the toroidal ISM field appears as circular lines inside the galaxy. Here, the wrapped ICM field is squeezed at $z = \pm R_{200}, x = 0$, and diverges at $x = \pm R_{200}, z = 0$.

\section{Results}
\label{sec:results}
We summarize the parameter space explored by our simulations of anisotropic thermal conduction on the evolution of the ISM in an ICM wind in
Table~\ref{tab:simulations}.

\begin{figure*}
\begin{deluxetable*}{lccc}
\tablecaption{Summary of simulations \label{tab:simulations}}
\tablehead{\colhead{Simulation} & \colhead{Magnetic field orientation} & 
\colhead{Magnetic field morphology} & \colhead{Anisotropic thermal conduction?} \\ 
\colhead{} & \colhead{($\parallel$ or $\perp$)} & \colhead{(continuous / shielded)} & \colhead{}} 
\startdata
cont-nocond-par &  $\mathbf{B} \parallel \mathbf{v_{\rm ICM}}$ &  continuous $\mathbf{B}$ field &  No \\
cont-nocond-perp &  $\mathbf{B} \perp \mathbf{v_{\rm ICM}}$ &  continuous $\mathbf{B}$ field &  No  \\
cont-cond-par &  $\mathbf{B} \parallel \mathbf{v_{\rm ICM}}$ &  continuous $\mathbf{B}$ field &  Yes \\
cont-cond-perp &  $\mathbf{B} \perp \mathbf{v_{\rm ICM}}$ &  continuous $\mathbf{B}$ field &  Yes \\
shield-nocond-par &  $\mathbf{B} \parallel \mathbf{v_{\rm ICM}}$ &  shielded $\mathbf{B}$ field &  No \\
shield-nocond-perp &  $\mathbf{B} \perp \mathbf{v_{\rm ICM}}$ &  shielded $\mathbf{B}$ field &  No  \\
shield-cond-par &  $\mathbf{B} \parallel \mathbf{v_{\rm ICM}}$ &  shielded $\mathbf{B}$ field &  Yes \\
shield-cond-perp &  $\mathbf{B} \perp \mathbf{v_{\rm ICM}}$ &  shielded $\mathbf{B}$ field &  Yes \\
\enddata
\end{deluxetable*}
\end{figure*}

\subsection{Magnetized ICM and ISM with Continuous Magnetic Fields, No Conduction }
\label{sec:nocond_cont}

\begin{figure*}[!htbp]
  \begin{center}
    {\includegraphics[width=\textwidth]{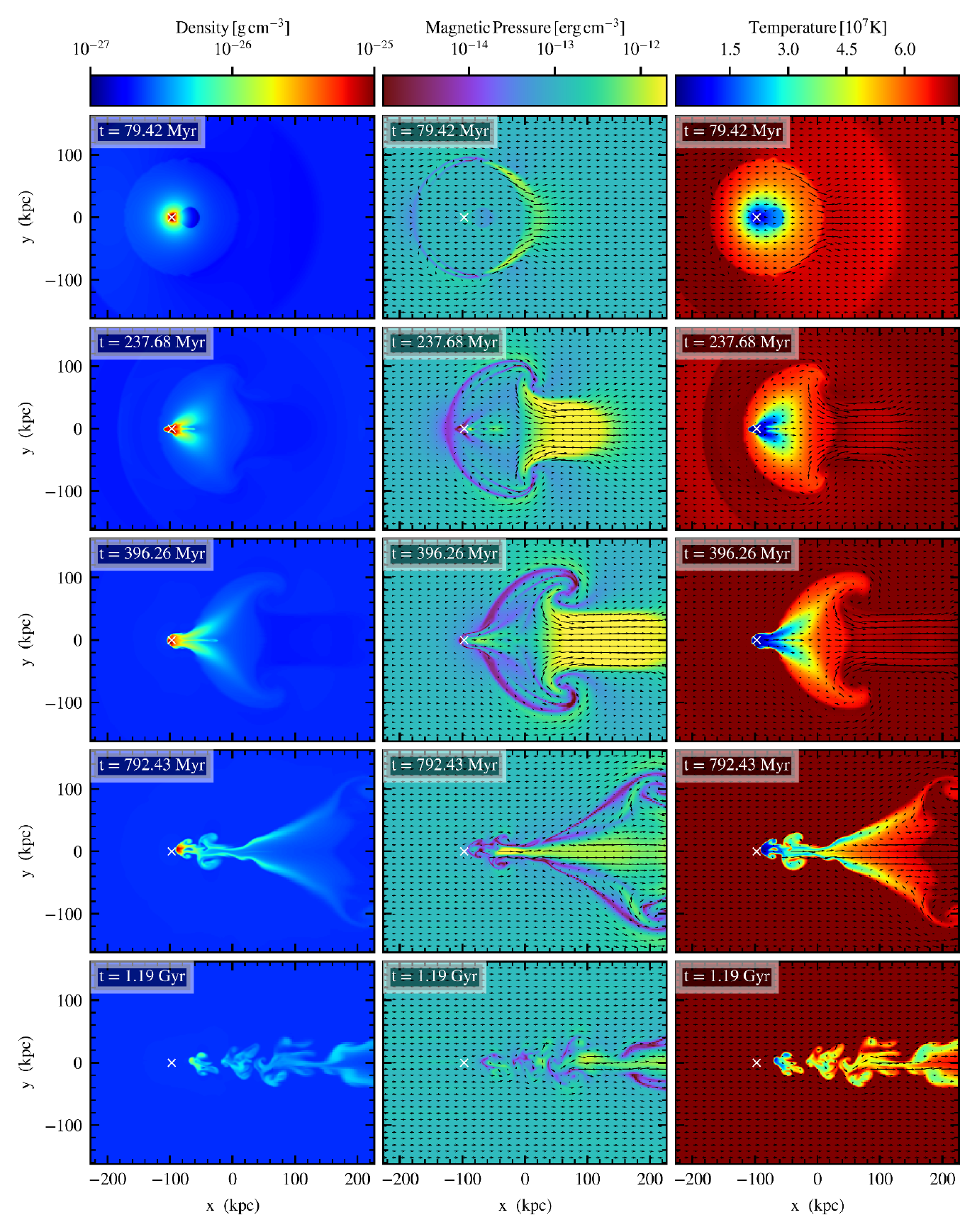}}
     \caption{Slices of gas density, magnetic pressure, and temperature in $\mathbf{B} \parallel \mathbf{v_{\rm ICM}}$ simulation without thermal conduction at $t = 80$ Myr, $t = 238$ Myr, $t = 400$ Myr, $t = 800$ Myr, and $t = 1200$ Myr. The white crosses mark the center of the galaxy. An animation for this figure is available. \label{fig:bxnocond}}
  \end{center}  
\end{figure*}

\begin{figure*}[!htbp]
  \begin{center}
    {\includegraphics[width=\textwidth]{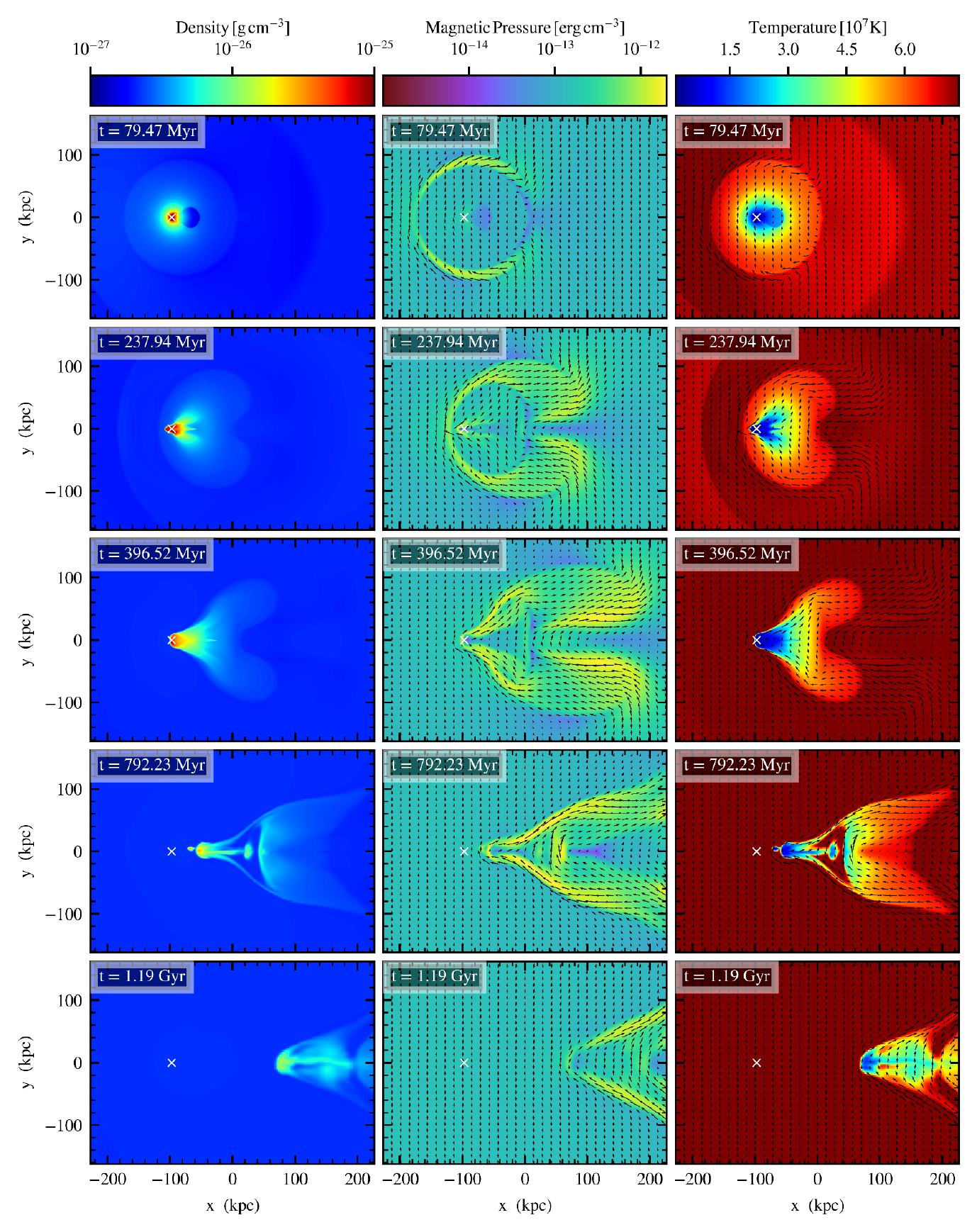}}
     \caption{Slices of gas density, magnetic pressure, and temperature in $\mathbf{B} \perp \mathbf{v_{\rm ICM}}$ simulation without thermal conduction at $t = 80$ Myr, $t = 238$ Myr, $t = 400$ Myr, $t = 800$ Myr, and $t = 1200$ Myr. An animation for this figure is available.\label{fig:bynocond}}
  \end{center}  
\end{figure*}

In our first series of simulations (cont-nocond-par and cont-nocond-perp), a uniform, continuous magnetic field threads the ICM and ISM. We use two extreme magnetic field orientations: (1) $\mathbf{B} \parallel \mathbf{v_{\rm ICM}}$, along the $x$-axis of the simulation box, and (2) $\mathbf{B} \perp \mathbf{v_{\rm ICM}}$. These simulations demonstrate the behavior of the magnetic field as it is swept around the galaxy, by the ICM wind in the frame of the galaxy and by the galaxy's orbital motion in the frame of the ICM. Using these simulations, we also study the effect of magnetic field orientation on ram-pressure-stripped ISM gas. Snapshots from these simulations are shown in Figures~\ref{fig:bxnocond} and \ref{fig:bynocond},  where the three columns show the evolution in density, magnetic pressure ($P_{\rm B} = \mathbf{B}^2 / 8 \pi$), and temperature in the $z = 0$ slice of the simulation box. The magnetic pressure and temperature slices are annotated with arrows showing the direction of the magnetic field.

Qualitative differences in the evolution of the hydrodynamic components as well as the magnetic field structure in the two extreme field orientations are immediately evident. At early times ($t \lesssim 240$ Myr) in the $\mathbf{B} \parallel \mathbf{v_{\rm ICM}}$ case, seen in Figure~\ref{fig:bxnocond}, the field at the leading edge of the galaxy diverges as it is advected along with the wind around the surface of the galaxy, and the magnetic pressure drops at the galaxy surface. In contrast, when $\mathbf{B} \perp \mathbf{v_{\rm ICM}}$, seen in Figure~\ref{fig:bynocond}, the field is compressed at the galaxy-ICM interface, and the magnetic pressure increases at the galaxy's surface. This is because field lines are compressed by the wind, resulting in increased magnetic field strength at the galaxy's leading edge, a phenomenon referred to as `magnetic draping' \citep{Lyutikov06,Dursi08}.  At later times ($240 $ Myr $\lesssim t \lesssim 800$ Myr), the magnetic field is enhanced in the stripped ISM tail and the ICM behind the galaxy; the morphology of these enhanced magnetic field regions are also distinct in the two configurations. When $\mathbf{B} \parallel \mathbf{v_{\rm ICM}}$, the field behind the galaxy is stretched in the direction of the ICM wind, resulting in an increase in magnetic pressure. This region does not correspond to the stripped ISM gas; there is no corresponding density enhancement. In fact, there is a decrease in density and thermal pressure in this magnetized wake; the total pressure ($P_{\rm thermal} + P_{\rm B}$) remains nearly constant in the magnetized wake and in the surrounding ICM. When $\mathbf{B} \perp \mathbf{v_{\rm ICM}}$, the enhancement in magnetic pressure behind the galaxy is not as strong, except in localized regions where the field is swept around the ISM component. When the ISM is almost completely stripped by ram pressure ($t \gtrsim 1$ Gyr), the ICM magnetic field largely returns to it's initial configuration except in those regions with remnant ISM gas. In the $\mathbf{B} \parallel \mathbf{v_{\rm ICM}}$ case, the stripped ISM tail becomes turbulent, corresponding to magnetic field reversals and alternating high and low magnetic pressure. In the $\mathbf{B} \perp \mathbf{v_{\rm ICM}}$ case, the stripped remnant ISM tail continues to lose gas, but is not turbulent, and is surrounded by a smooth shield of increased magnetic pressure.

The corresponding evolution in density and temperature structures of the stripped ISM are distinct, particularly at late times ($t \gtrsim 400$ Myr). Stripped ISM tails, seen in their density and temperature slices, are long and narrow when $\mathbf{B} \parallel \mathbf{v_{\rm ICM}}$; when $\mathbf{B} \perp \mathbf{v_{\rm ICM}}$ the tail is wider and smoother. There is no turbulent disruption of the tail in the $\mathbf{B} \perp \mathbf{v_{\rm ICM}}$  case as the stripped gas is confined within a smooth region whose outer surface is defined by a shell of amplified of magnetic field strength. Although the magnetic field orientation qualitatively affects the structure of the stripped tail, the overall amount of gas stripped does not change significantly; most of the ISM is stripped by $t \gtrsim 1$ Gyr. In both field configurations, the presence of the magnetic field suppresses the formation of Kelvin-Helmholtz instabilities on the sides of the galaxy, when compared to the identical simulation in Paper I without magnetic fields.  

\subsection{Parallel and perpendicular magnetic fields, anisotropic conduction }
\label{sec:cond_cont}

\begin{figure*}[!htbp]
  \begin{center}
    {\includegraphics[width=\textwidth]{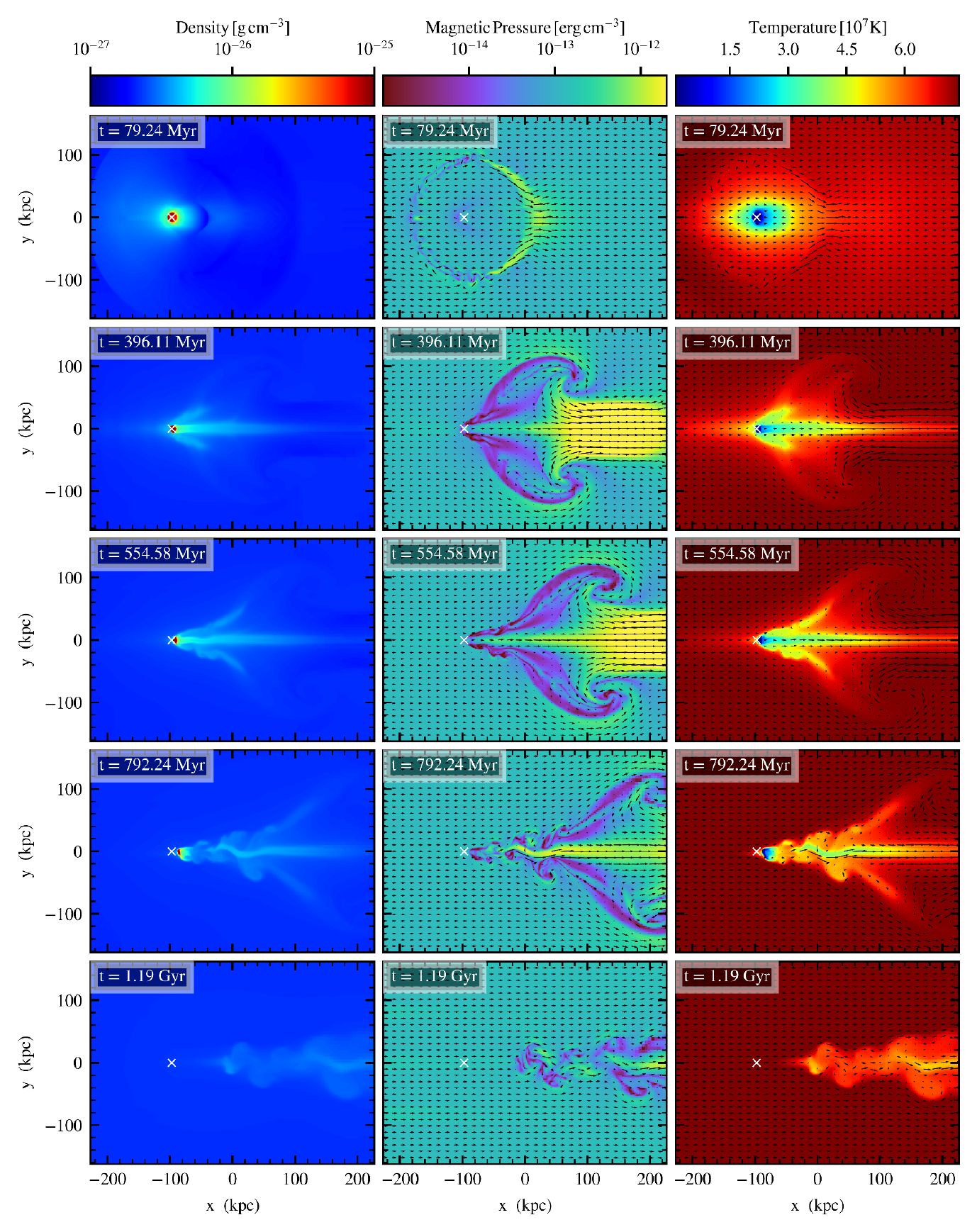}}
     \caption{Slices of gas density, magnetic pressure, and temperature in $\mathbf{B} \parallel \mathbf{v_{\rm ICM}}$ simulation with a connected ICM + ISM magnetic field and anisotropic thermal conduction at $t = 80$ Myr, $t = 238$ Myr, $t = 400$ Myr, $t = 800$ Myr, and $t = 1200$ Myr. An animation for this figure is available. \label{fig:bxcond}}
  \end{center}  
\end{figure*}

\begin{figure*}[!htbp]
  \begin{center}
    {\includegraphics[width=\textwidth]{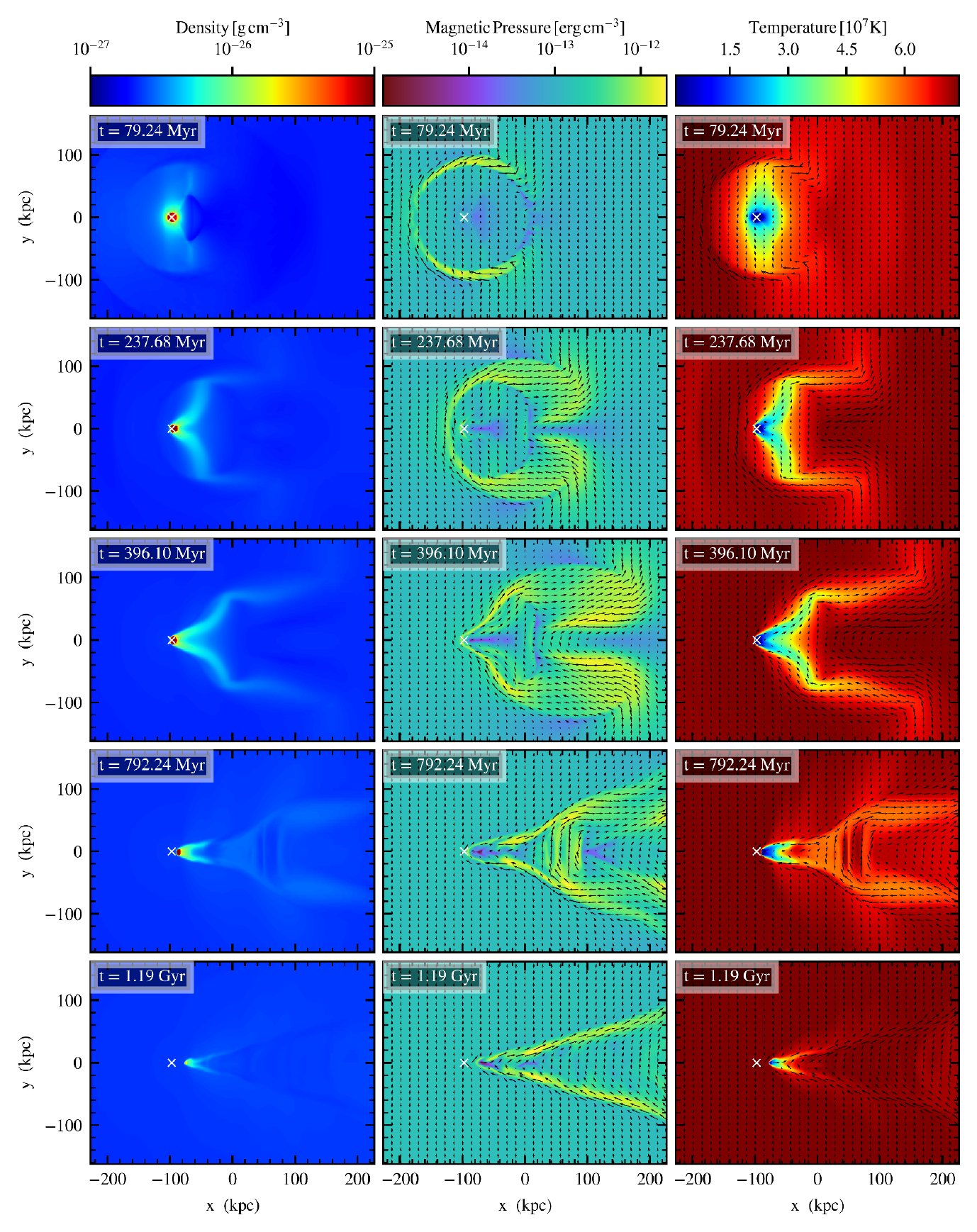}}
     \caption{Slices of gas density, magnetic pressure, and temperature in $\mathbf{B} \perp \mathbf{v_{\rm ICM}}$ simulation with a connected ICM + ISM magnetic field and anisotropic thermal conduction at $t = 80$ Myr, $t = 238$ Myr, $t = 400$ Myr, $t = 800$ Myr, and $t = 1200$ Myr. An animation for this figure is available. \label{fig:bycond}}
  \end{center}  
\end{figure*}

In our second series of simulations (cont-cond-par and cont-cond-perp), we include the effects of anisotropic thermal conduction on the $2.8 \times 10^{11} \msun$ galaxy being ram pressure stripped, with a continuous magnetic field between the ICM and ISM. With the magnetic field being fully continuous through the ICM and ISM, a somewhat unrealistic scenario, heat can flow efficiently from the ICM to the ICM thereby potentially overestimating the effectiveness of anisotropic thermal conduction in real cluster galaxies. In contrast, we also run a series of simulations where the magnetic field in the ICM and ISM are completely disjoint; these simulations may underestimate the real effectiveness of anisotropic conduction. Between these two extreme orientations, we therefore likely capture the range of effectiveness of anisotropic conduction. 

Slices of density, magnetic pressure, and temperature in the cont-cond-par and cont-cond-perp simulations are seen in Figures~\ref{fig:bxcond} and \ref{fig:bycond}. Comparing these snapshots, in particular the density and temperature slices, with the corresponding snapshots in cont-nocond-par and cont-nocond-perp simulations (Figures~\ref{fig:bxnocond} and \ref{fig:bynocond}), the flow of cool dense gas along magnetic field lines is immediately apparent. We find in Paper I that in the presence of isotropic saturated conduction, the transfer of heat from the hot ICM to the cool ISM leads to the expansion and heating of the ISM, which evaporates in $t = 160$ Myr. When thermal conduction is anisotropic, heat is transferred along the direction of the local magnetic field. This results in a flow of cooler, denser gas along the ICM magnetic field, albeit at a significantly slower rate than the rapid expansion and evaporation due to isotropic thermal conduction. 

When $\mathbf{B} \parallel \mathbf{v_{\rm ICM}}$ (Figure~\ref{fig:bxcond}), the anisotropic evaporation of cool, dense gas is primarily parallel to the $x$-axis, where the magnetic field starts to diverge around the surface of the galaxy and converges in a coherent tail behind the galaxy. This feature is visible as additional filaments of gas (compared to the slices in Figure~\ref{fig:bxnocond}) ahead of and behind the galaxy. These filaments narrow and elongate with time before eventually dissipating as the gas is heated to the temperature of the surrounding ICM. The rate at which ISM gas evaporates (and is ram pressure stripped) is moderately faster compared to the case without anisotropic conduction, but significantly slower than when conduction is fully isotropic. At late times when the stripped, rapidly diminishing tail becomes turbulent, conduction ensures that heat flows through and within the stripped tail, smearing out jumps and discontinuities in density and temperature.

In the $\mathbf{B} \parallel \mathbf{v_{\rm ICM}}$ case, the direction along which the magnetic field is advected with the ICM wind is parallel to the direction of heat flow; therefore, the filaments of cool dense gas grow and eventually dissipate along the same direction. When $\mathbf{B} \perp \mathbf{v_{\rm ICM}}$ (Figure~\ref{fig:bycond}), the magnetic field wraps around the galaxy and is advected away from the galaxy by the ICM wind. This affects the formation and evolution of evaporating ISM filaments. Initially ($t \simeq 80$ Myr), the density and temperature slices of the ISM appear elongated along the $y$-axis; when $\mathbf{B} \parallel \mathbf{v_{\rm ICM}}$ this elongation is along the $x$-axis. The magnetic field around the outer surface of the galaxy is advected and amplified in the direction of the ICM wind, resulting in a field configuration that bends by $\sim$90\arcdeg\ in the region where the ICM wind must converge behind the galaxy. Heat flows from the ICM to the ISM primarily through this region where the ICM wind converges, forming two perpendicular bent filaments of cool, dense gas that flow away from the galaxy. 

\begin{figure*}[!htbp]
  \begin{center}
    \subfigure[Density]
    {\includegraphics[width=0.32\textwidth]{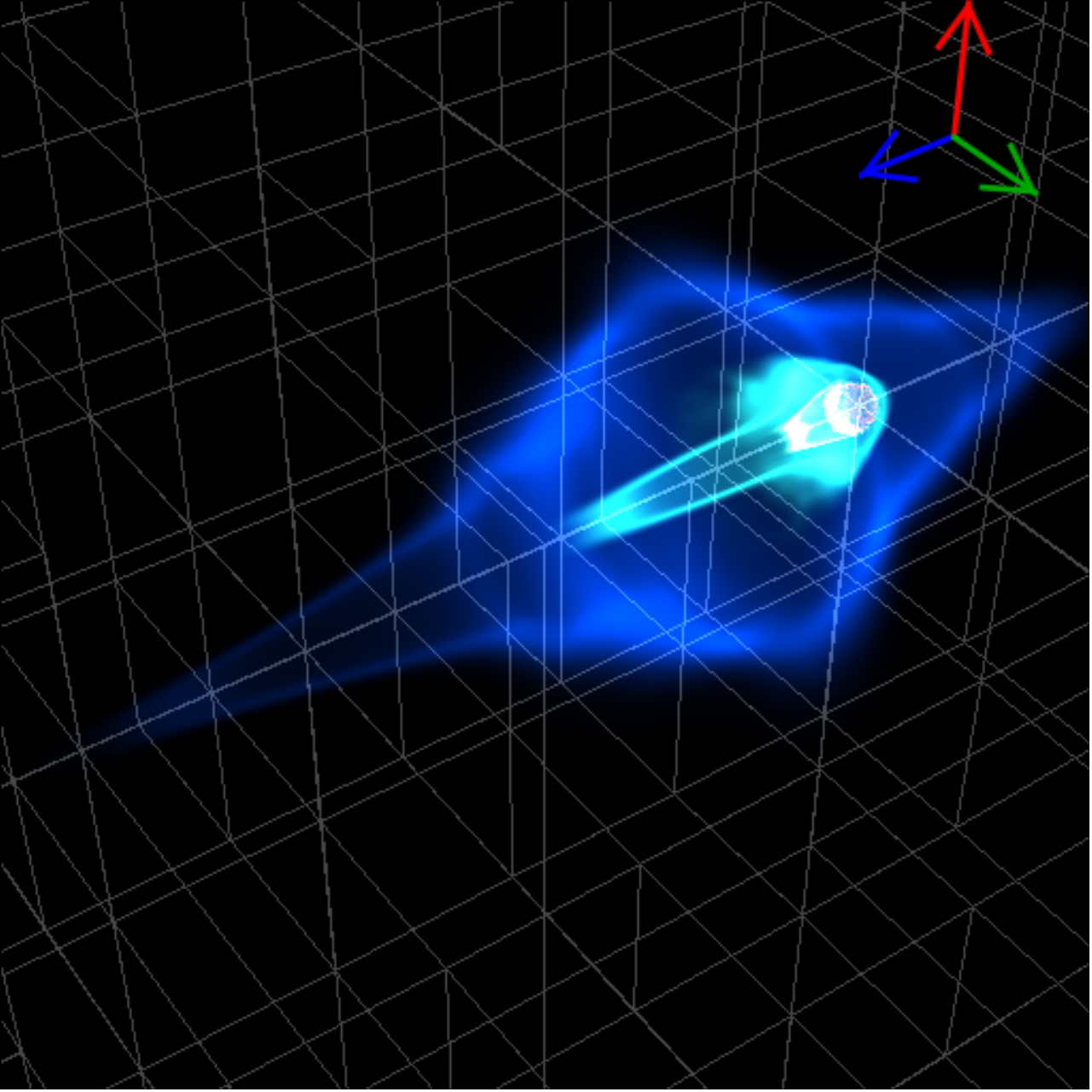}}
    \subfigure[Magnetic Pressure]
    {\includegraphics[width=0.32\textwidth]{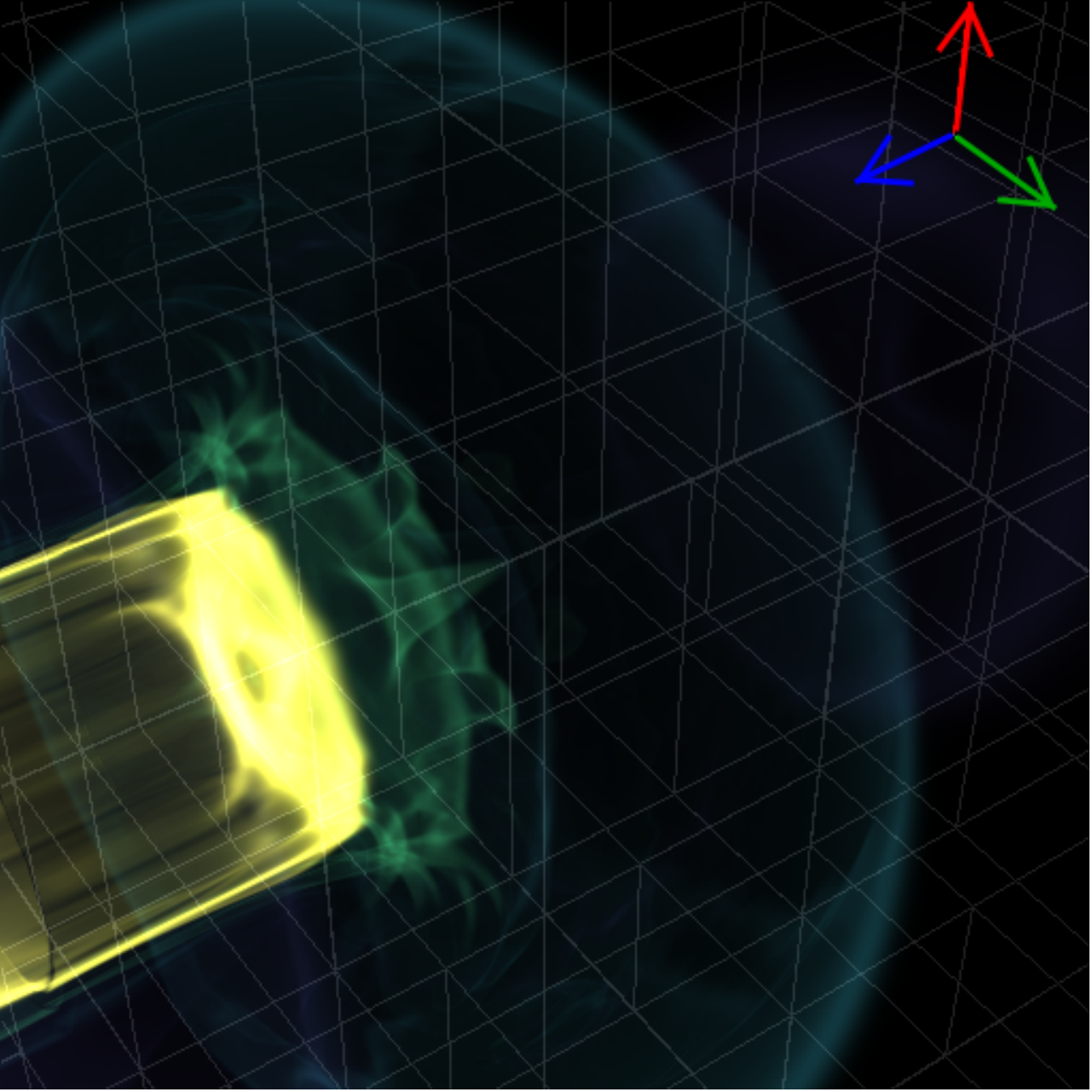}}
    \subfigure[Temperature]
    {\includegraphics[width=0.32\textwidth]{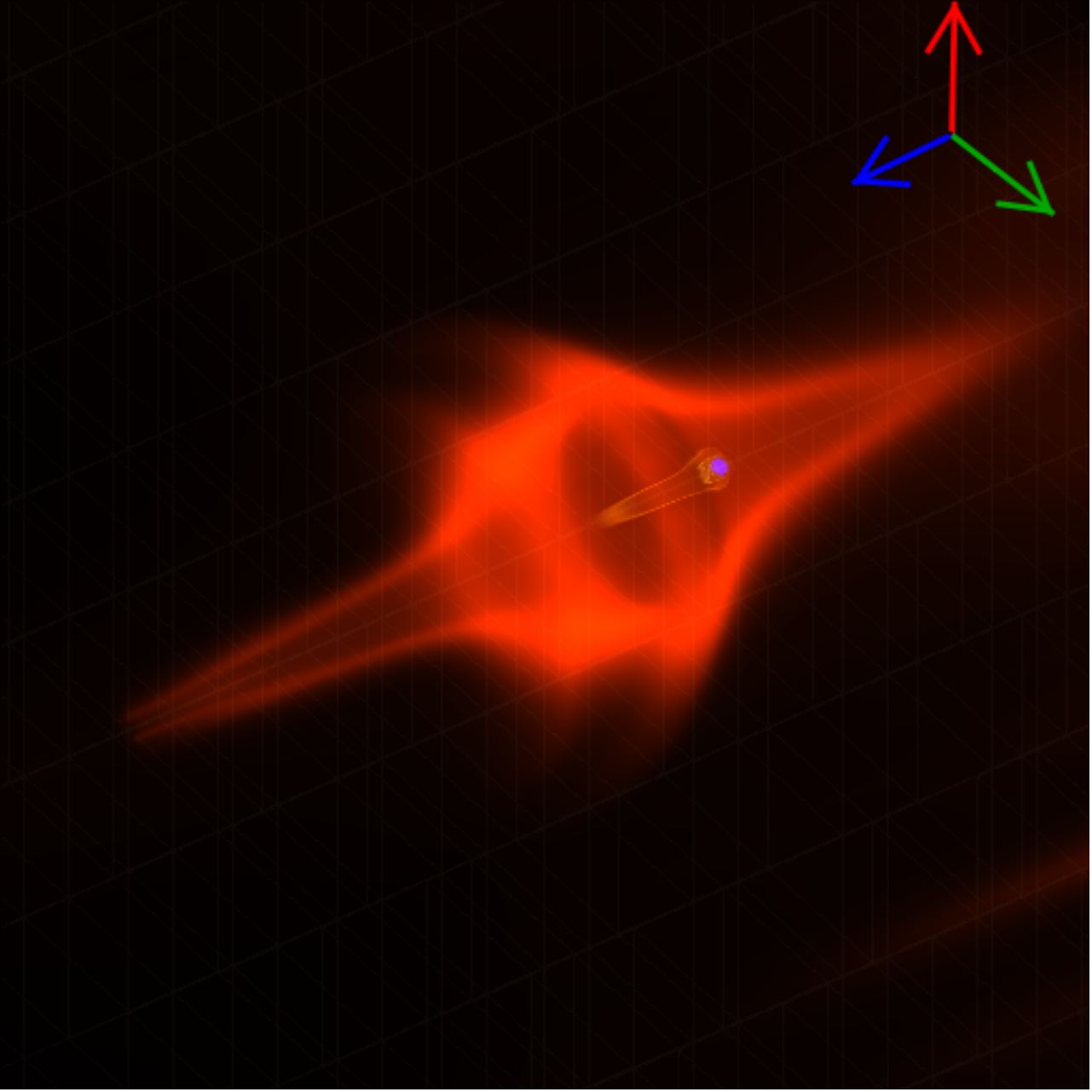}}
     \caption{Three dimensional volume renderings of density, magnetic pressure, and temperature isocontours at $t = 400$ Myr in the  $\mathbf{B} \parallel \mathbf{v_{\rm ICM}}$ simulation with a connected ICM + ISM magnetic field and anisotropic thermal conduction (simulation cont-cond-par). The upper right hand corner shows the directions of the three axes, where the blue line is the direction of the $x$-axis, the red line is the direction of the $y$-axis, and the green line is the direction of the $z$-axis. Grid lines are overdrawn at the lowest resolution level. The center of the dark matter distribution of the galaxy is at the grid vertex associated with the highest gas density.    
     \label{fig:3d_bxnocond}}
  \end{center}  
\end{figure*}

\begin{figure*}[!htbp]
  \begin{center}
    \subfigure[Density]
    {\includegraphics[width=0.32\textwidth]{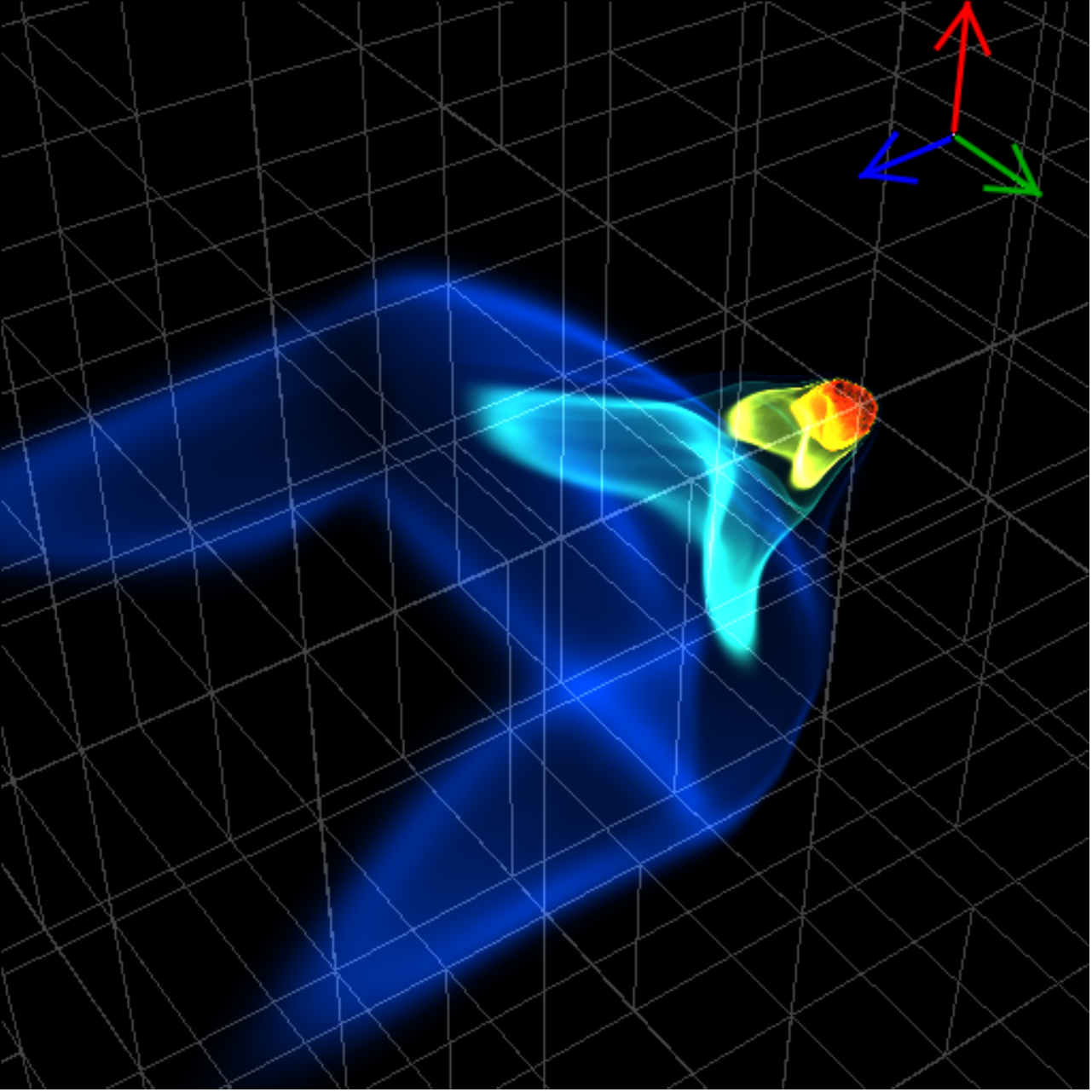}}
    \subfigure[Magnetic Pressure]
    {\includegraphics[width=0.32\textwidth]{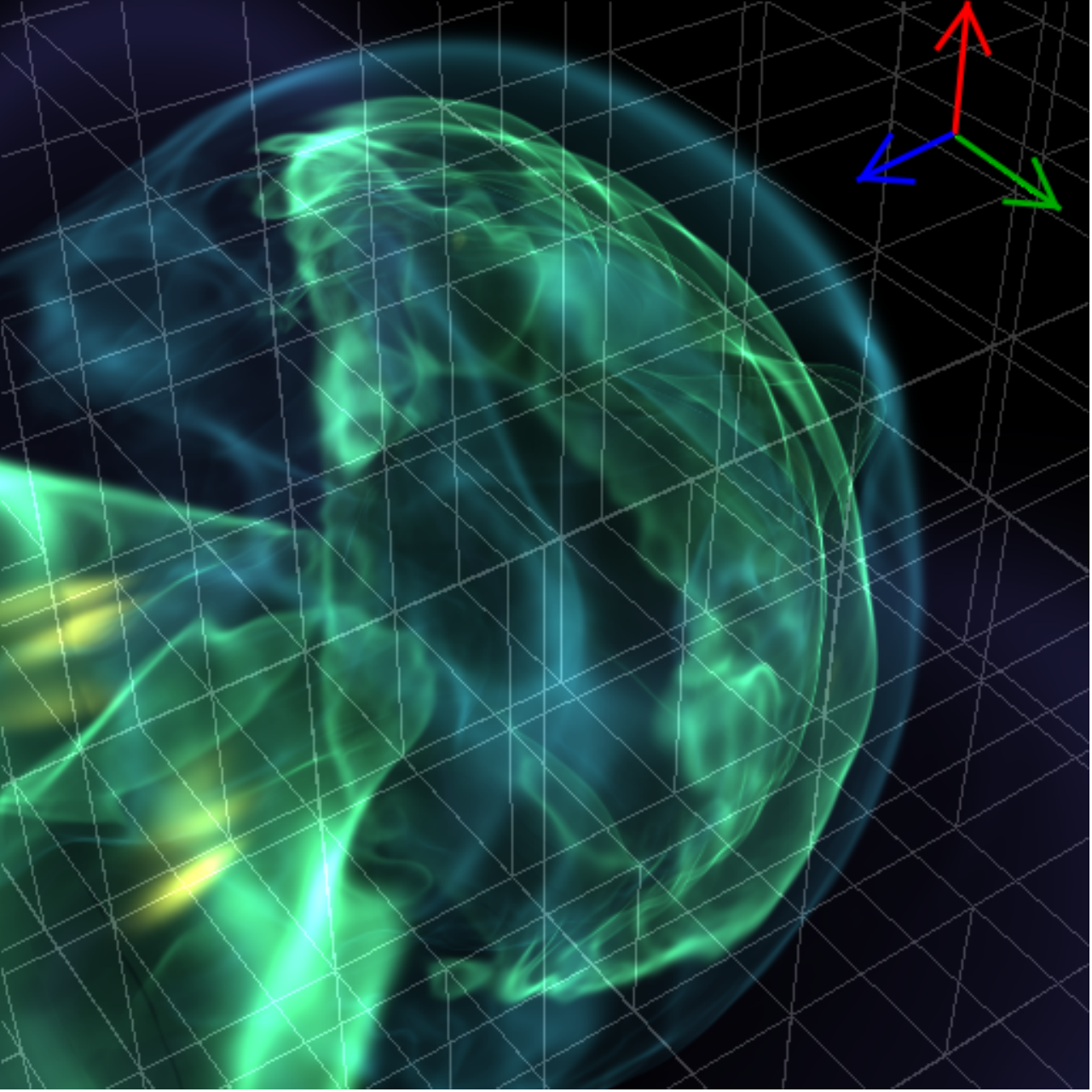}}
    \subfigure[Temperature]
    {\includegraphics[width=0.32\textwidth]{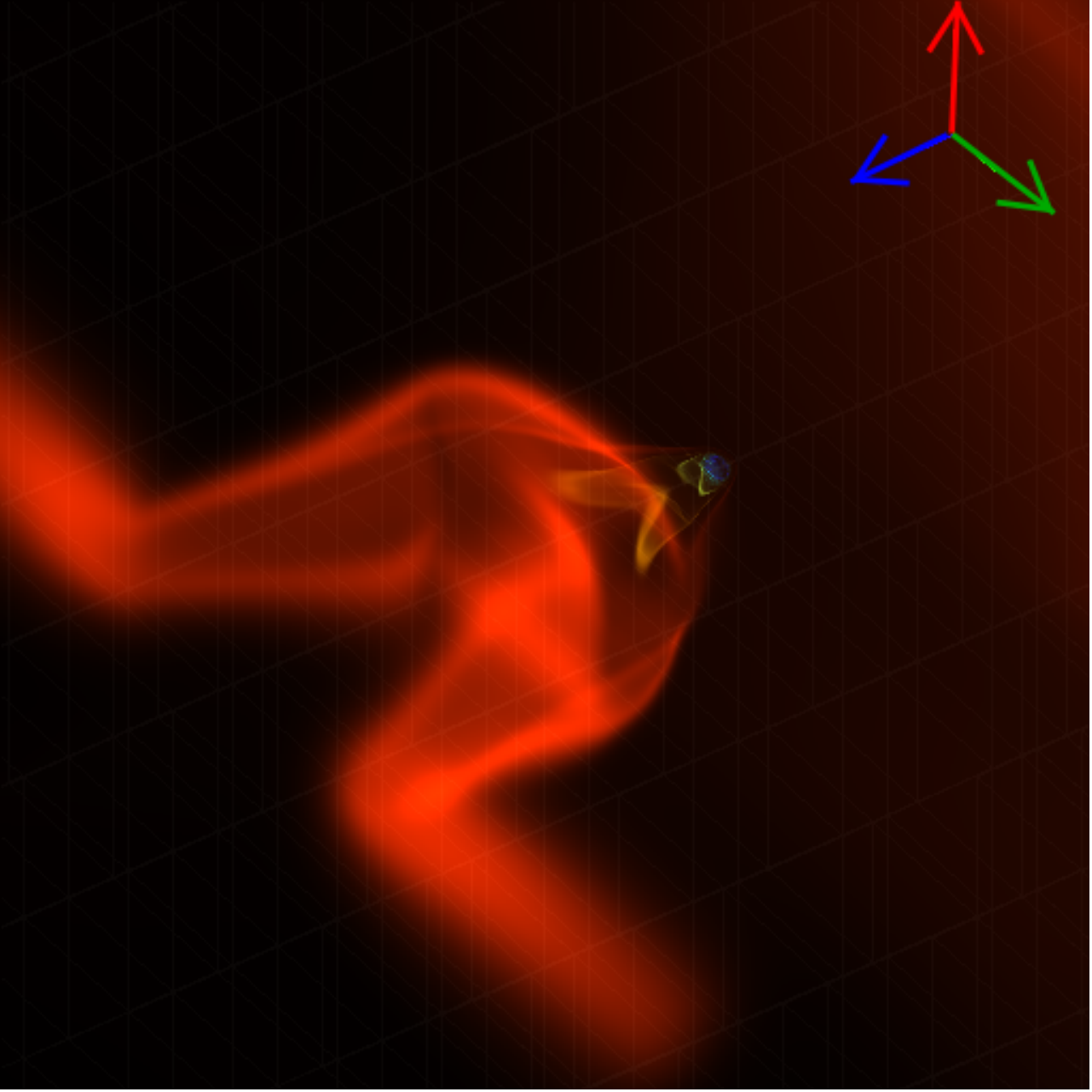}}
     \caption{Three dimensional volume renderings of density, magnetic pressure, and temperature at $t = 400$ Myr in the $\mathbf{B} \perp \mathbf{v_{\rm ICM}}$ simulation with a connected ICM + ISM magnetic field and anisotropic thermal conduction (simulation cont-cond-perp). The upper right hand corner shows the directions of the three axes, where the blue line is the direction of the $x$-axis, the red line is the direction of the $y$-axis, and the green line is the direction of the $z$-axis. Grid lines are overdrawn at the lowest resolution level. The center of the dark matter distribution of the galaxy is at the grid vertex associated with the highest gas density.       
      \label{fig:3d_bynocond}}
  \end{center}  
\end{figure*}

These features are highlighted in Figures~\ref{fig:3d_bxnocond} and \ref{fig:3d_bynocond}, which are 3D volume renderings of density, magnetic pressure, and temperature isocontours at $t = 400$ Myr. When $\mathbf{B} \parallel \mathbf{v_{\rm ICM}}$  (Fig.~\ref{fig:3d_bxnocond}), we see that the density and temperature isocontours, which trace ISM gas, are in the same direction as the strongly magnetized region behind the galaxy; there is also a region of ISM gas ahead of the galaxy in the direction of the local magnetic field. When $\mathbf{B} \perp \mathbf{v_{\rm ICM}}$  (Fig.~\ref{fig:3d_bynocond}), the ICM magnetic field in the $y$ direction and the advection of the magnetic field by the wind in the $x$ direction  forms filaments bent by $\sim$90 degrees, highlighted in the density and temperature isocontours. These features are not axially symmetric about the x-axis, but are mainly confined to the x-y plane (the plane $z = 0$).

The magnetic pressure isocontours highlight the regions where the field is amplified in both cases. When $\mathbf{B} \parallel \mathbf{v_{\rm ICM}}$, the magnetic field ahead of the galaxy diverges, and the field behind the galaxy is strongly amplified (central panel, Fig.~\ref{fig:3d_bxnocond}). In the $\mathbf{B} \perp \mathbf{v_{\rm ICM}}$  case, the field is compressed and amplified ahead of the galaxy.

At the galaxy's leading edge, the component of the local magnetic field perpendicular to the direction of the wind is amplified and wrapped around the surface of the galaxy, effectively shielding it by suppressing thermal conduction across the wrapped magnetic field. At late times, when most of the ISM has been stripped by ram pressure, the shielding effect of the magnetic field on the galaxy's leading edge decreases. Heat flows to the ISM gas in the galaxy's core. Eventually, the ISM evaporates and is stripped by $t \simeq 1200$ Myr. Notably, as in the case without thermal conduction, the stripped ISM tail does not become turbulent when $\mathbf{B} \perp \mathbf{v_{\rm ICM}}$. Although the morphology and structure of the stripped tails and filaments of evaporation are remarkably distinct for the two magnetic field orientations explored, the overall rates of gas loss, while higher than in the absence of conduction, are similar in both cases.

In the presence of anisotropic conduction, the total gas mass loss time is $t \simeq 1200 $ Myr, comparable to the ram pressure stripping time. The isotropic evaporation timescale for the same galaxy is $t \simeq 160 $ Myr. Ram pressure stripping therefore plays a dynamically significant role in removing gas when thermal conduction is anisotropic, compared to the isotropic case when gas rapidly evaporates before ram pressure can remove a significant fraction of the ICM. Although anisotropic thermal conduction has a significant effect on the hydrodynamic properties of the ISM and ICM, the magnetic field is largely unaffected, as apparent when compared to the case without conduction in \S~\ref{sec:nocond_cont}.

\subsection{Shielded galaxy, parallel and perpendicular magnetic fields, no conduction }
\label{sec:nocond_shield}

\begin{figure*}[!htbp]
  \begin{center}
    {\includegraphics[width=\textwidth]{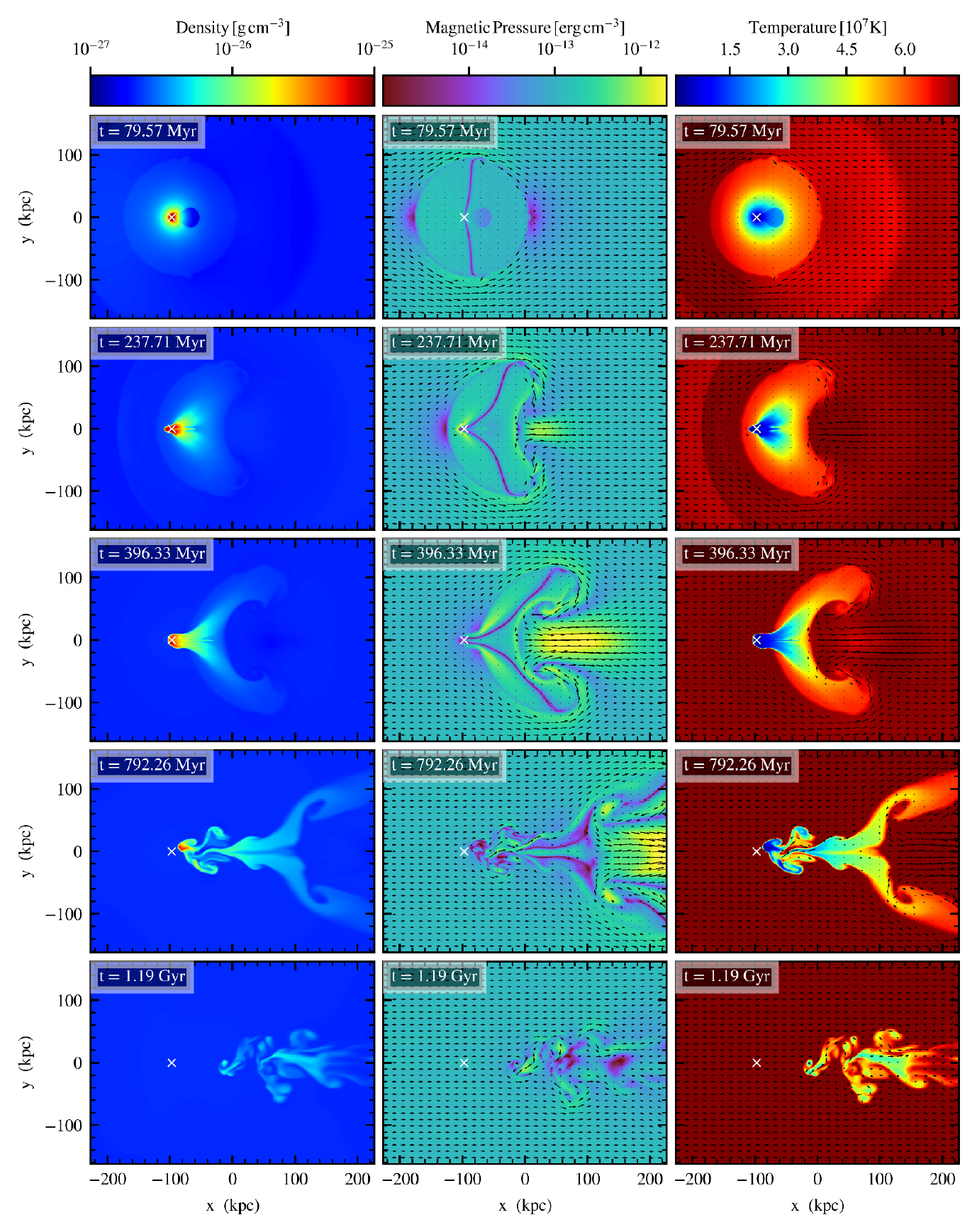}}
     \caption{Slices of gas density, magnetic pressure, and temperature in $\mathbf{B} \parallel \mathbf{v_{\rm ICM}}$ simulation, with a shielded ICM field and toroidal ISM field, without thermal conduction at $t = 80$ Myr, $t = 238$ Myr, $t = 400$ Myr, $t = 800$ Myr, and $t = 1200$ Myr. An animation for this figure is available.\label{fig:bxnocondshield}}
  \end{center}  
\end{figure*}

\begin{figure*}[!htbp]
  \begin{center}
    {\includegraphics[width=\textwidth]{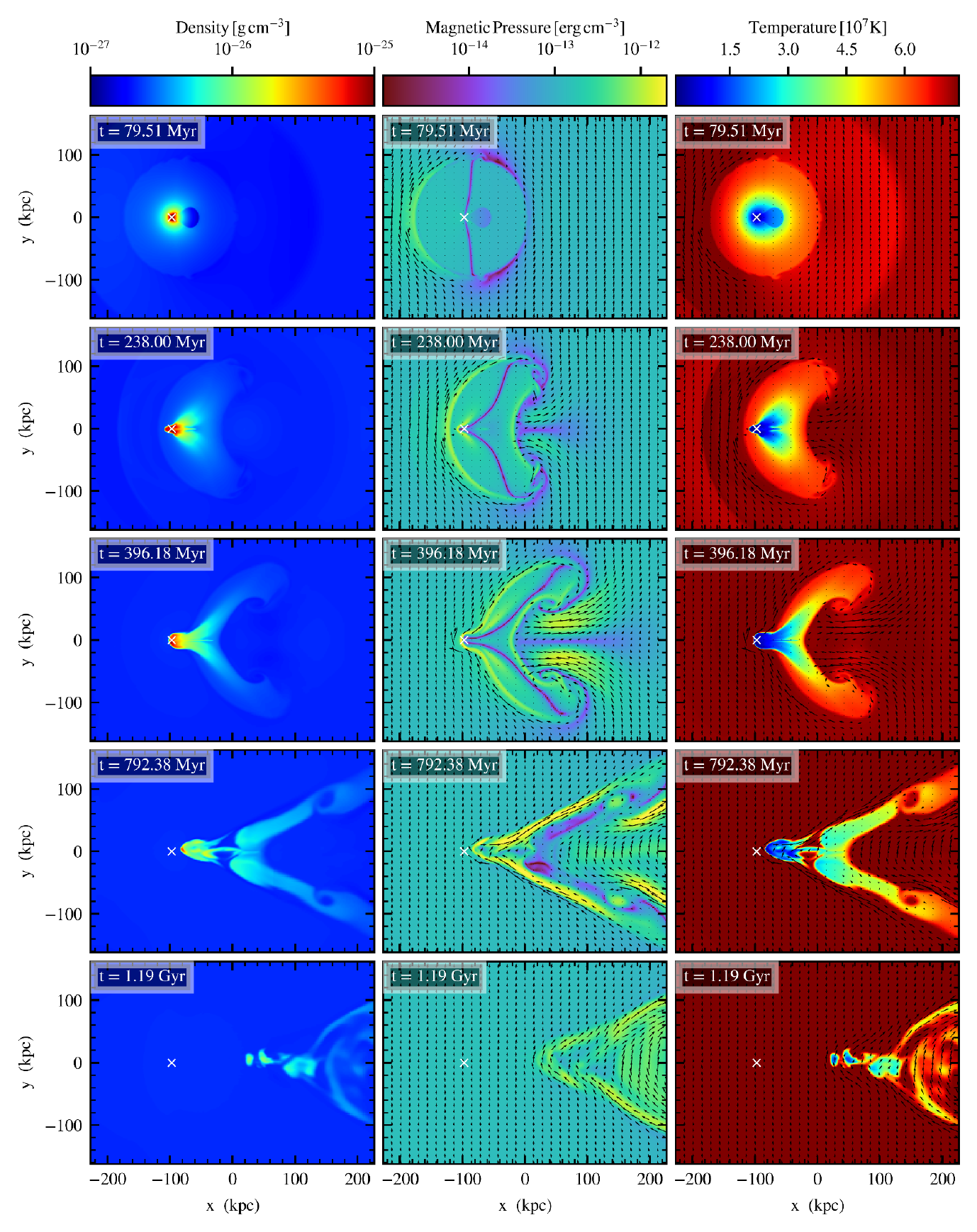}}
     \caption{Slices of gas density, magnetic pressure, and temperature in $\mathbf{B} \perp \mathbf{v_{\rm ICM}}$ simulation, with a shielded ICM field and toroidal ISM field, without thermal conduction at $t = 80$ Myr, $t = 238$ Myr, $t = 400$ Myr, $t = 800$ Myr, and $t = 1200$ Myr. An animation for this figure is available. \label{fig:bynocondshield}}
  \end{center}  
\end{figure*}

In our next series of simulations (shield-nocond-par and shield-nocond-perp) the ICM magnetic field is initially wrapped around the galaxy; the galaxy's magnetic field is toroidal in the $x-z$ plane. The results of our simulations are not sensitive to the plane of the galaxy's initial toroidal field  -- we performed a lower resolution simulation series with the galaxy's internal toroidal magnetic field in the $y-z$ plane, and did not find any significant differences in the evolution of the galaxy and its stripped tail. The galaxy's magnetic field is completely disjoint from the ICM magnetic field, an extreme and idealized opposing scenario in contrast to the previous experiments (cont-nocond- and cont-cond- in \S~\ref{sec:nocond_cont} and \S~\ref{sec:cond_cont}) where the magnetic field is continuous and initially independent of the local ISM and ICM distribution. With these two extreme ISM-ICM magnetic field orientations, we quantify the range of efficiency of anisotropic thermal conduction in evaporating gas. When the ISM and ICM magnetic fields are not connected, the flow of heat along magnetic field lines is severely restricted, and ideally, minimally efficient. Of course, with the bending, distortion, and re-alignment of magnetic field lines with the ICM wind, anisotropic conduction does not remain either  maximally or minimally efficient in either configuration. 

Slices of density, magnetic pressure, and temperature are in Figure~\ref{fig:bxnocondshield} (shield-nocond-par) and Figure~\ref{fig:bynocondshield} (shield-nocond-perp). We first compare the shielded $\mathbf{B} \parallel \mathbf{v_{\rm ICM}}$ case, to the continuous magnetic field case (\S~\ref{sec:nocond_cont} and Figure~\ref{fig:bxnocond}). Prominent features and differences visible initially ($t = 80$ Myr) are: (1) the magnetic field lines inside the galaxy, being toroidal in the $x-z$ plane, which is perpendicular to the plane of the page, are simply visible as dots showing magnetic field lines coming out of the page, (2) the central strip of low magnetic field strength corresponds to the central cylindrical region where the finite resolution of the grid results in a magnetic field strength close to zero, (3) the ICM magnetic field diverges at the equatorial plane and converges at the poles as it wraps around the galaxy. With the ICM magnetic field being initially wrapped around the galaxy, the flow of the ICM wind around the galaxy does not cause the magnetic field to bend to the same extent as in the cont-nocond-par case. Therefore, the layer of decreased field strength caused by a divergent magnetic field at $t = 240 - 400$ Myr on the galaxy's leading edge surface in cont-nocond-par is not prominent in shield-nocond-par beyond the initial divergence near the equatorial plane. The enhancement in field strength in the galaxy's wake is also less prominent in the shield-nocond-par model ($t = 400 - 800$ Myr). The ISM magnetic field, being disjoint from the ICM field, is not dragged along and amplified. At late times, the ram pressure stripped ISM tail continues to narrow and becomes turbulent, as in cont-nocond-par ($t = 800 - 1200$ Myr). The shielding effect of the ICM magnetic field on the ISM tail leads to more prominent `wings', as seen in the density and temperature slices at $t = 400 - 800$ Myr, since the shielded ISM does not fully occupy the region between the stripped ISM tail zones.

When $\mathbf{B} \perp \mathbf{v_{\rm ICM}}$ in shield-nocond-perp, the initial wrapped ICM field is also not amplified and modified as extensively as in the continuous field case in \S~\ref{sec:nocond_cont} and Figure~\ref{fig:bynocond}. The initial wrapped vertical ICM field is enhanced at the galaxy's magnetic equator (i.e., in the $y = 0$ plane) and diverges near the galaxy's magnetic poles (where $x = z = 0$, and $y = \pm R_{200}$). At $240 - 400$ Myr, ram pressure on the galaxy's leading edge compresses and amplifies the magnetic field. As ICM wind flows around the galaxy, the magnetic field around the galaxy is advected with the wind, becoming bent and amplified in the galaxy's wake. The shielded (disjoint) magnetic field configurations in the ISM and ICM result in a double-winged tail. While in the continuous $\mathbf{B} \perp \mathbf{v_{\rm ICM}}$ simulation, the stripped ISM tail has a cylindrical shell-like structure, the stripped tail here has two distinct wings, forming a double tailed structure ($t \simeq 400 - 800$ Myr). These features are further illustrated in Figure~\ref{fig:3d_doubletail} below.

When the ISM has been almost completely stripped by $t \simeq 1200$ Myr, the tail diffuses into the wake region.

A consequence of the disjoint ISM-ICM magnetic field configuration, both in the shield-nocond-par and shield-nocond-perp simulations, is that the density and temperature of the ISM evolve in a similar fashion up to $t \simeq 400$ Myr. The orientation of the external ICM magnetic field ($\mathbf{B} \parallel \mathbf{v_{\rm ICM}}$  or $\mathbf{B} \perp \mathbf{v_{\rm ICM}}$) does not affect the ISM at these early times. Only at later times, as the ISM is stripped away from the galaxy by ram pressure, does the ICM magnetic field orientation (shielded cone when $\mathbf{B} \perp \mathbf{v_{\rm ICM}}$, turbulent and tangled when $\mathbf{B} \parallel \mathbf{v_{\rm ICM}}$) affect the density and temperature distribution of the stripped ISM.
 
\subsection{Shielded galaxy, parallel and perpendicular magnetic fields, anisotropic conduction }
\label{sec:cond_shield}

\begin{figure*}[!htbp]
  \begin{center}
    {\includegraphics[width=\textwidth]{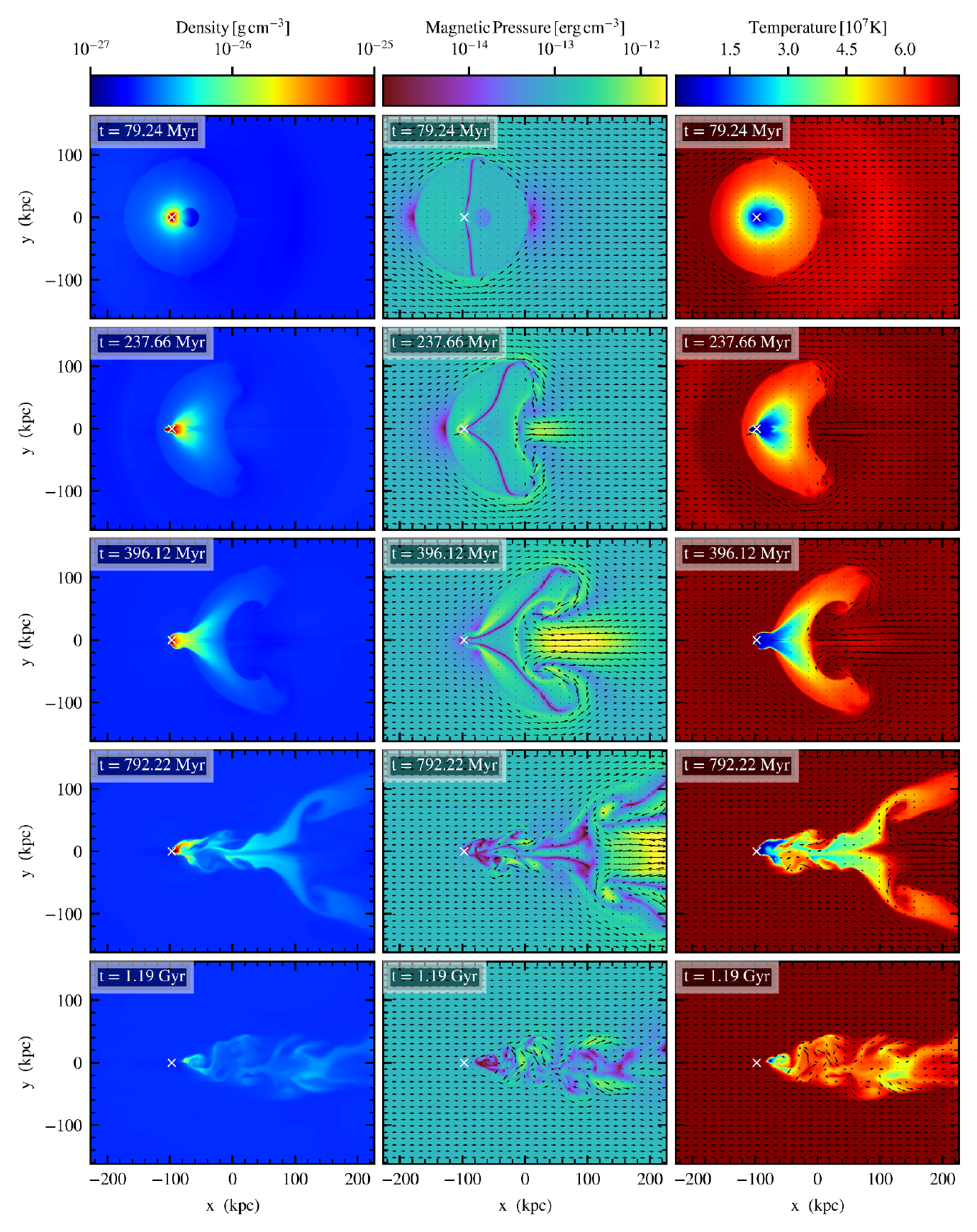}}
     \caption{Slices of gas  density, magnetic pressure, and temperature in $\mathbf{B} \parallel \mathbf{v_{\rm ICM}}$ simulation, with a shielded ICM field and toroidal ISM field, with anisotropic thermal conduction at $t = 80$ Myr, $t = 238$ Myr, $t = 400$ Myr, $t = 800$ Myr, and $t = 1200$ Myr. An animation for this figure is available.\label{fig:bxcondshield}}
  \end{center}  
\end{figure*}

\begin{figure*}[!htbp]
  \begin{center}
    {\includegraphics[width=\textwidth]{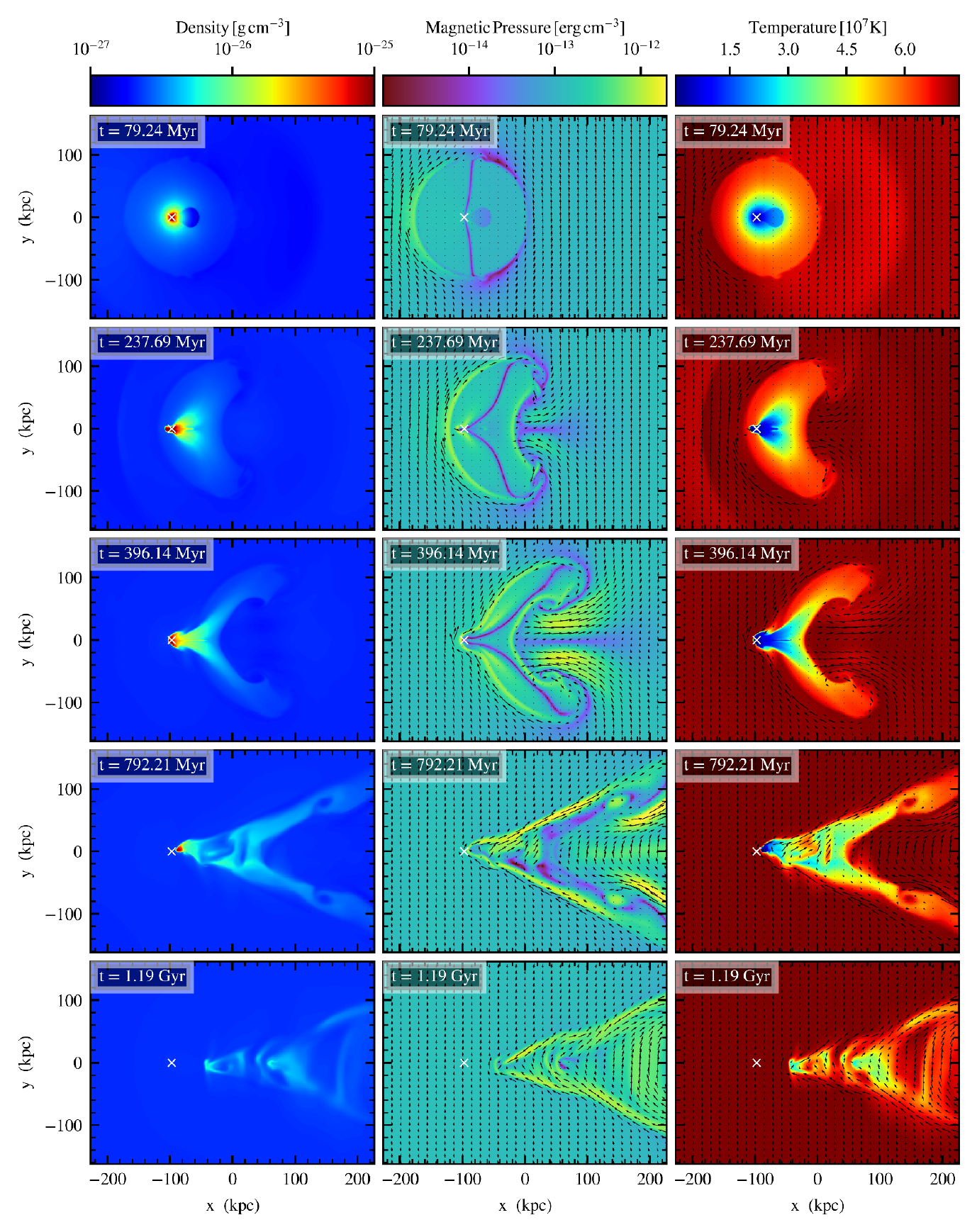}}
     \caption{Slices of gas density, magnetic pressure, and temperature in $\mathbf{B} \perp \mathbf{v_{\rm ICM}}$ simulation, with a shielded ICM field and toroidal ISM field, with anisotropic thermal conduction at $t = 80$ Myr, $t = 238$ Myr, $t = 400$ Myr, $t = 800$ Myr, and $t = 1200$ Myr. An animation for this figure is available. \label{fig:bycondshield}}
  \end{center}  
\end{figure*}
 
In our final simulation series (shield-cond-par and shield-cond-perp), we incorporate the effects of anisotropic thermal conduction with a disjoint ICM-ISM magnetic field configuration. The ISM's initial temperature distribution is isotropic, so regions at a given galaxy-centric radius are at the same temperature. Toroidal magnetic field lines in any given plane within the galaxy therefore traverse regions at the same temperature; i.e., the temperature gradient along magnetic field lines within the galaxy is zero. In the absence of a temperature gradient along the local magnetic field, heat does not initially flow within the galaxy due to anisotropic conduction. 

The shielding of the ICM magnetic field around the galaxy also suppresses the flow of heat from the ISM to the ICM. Even when the ISM and ICM magnetic fields are continuous (\S~\ref{sec:cond_cont}), the anisotropic nature of thermal conduction severely suppresses the flow of heat from the ICM to the ISM compared to the fully isotropic scenario in Paper I. When the ISM and ICM magnetic fields are initially connected, heat flows primarily along magnetic field lines from the ICM to within the galaxy, and evaporated gas flows in narrow, cool, dense filaments. When the ISM magnetic field is not initially connected to the ICM, as in this set of simulations, there are fewer pathways to transport heat between the ISM and ICM. 

The snapshots in Figures~\ref{fig:bxcondshield} and ~\ref{fig:bycondshield} show the evolution of density, magnetic pressure, and temperature in the $z = 0$ plane of the shield-cond-par and shield-cond-perp simulations. We first compare these results to the evolution of the galaxy in the cont-cond-par and cont-cond-perp simulations. When $\mathbf{B} \parallel \mathbf{v_{\rm ICM}}$ (Figures~\ref{fig:bxcond} and ~\ref{fig:bxcondshield}), in the shielded case, there is no cool dense filament of gas ahead of the galaxy's leading edge: the ISM magnetic field is not connected to the ICM magnetic field which diverges around the galaxy at $y = 0$.

However, in the galaxy's wake, there is some mixing of cool ISM gas due to the flow of heat from the ICM to the ISM at $t \simeq 240 - 400$ Myr, where the ICM field likely reconnects with the ISM field due to numerical diffusion and reconnection. Our simulations assume ideal MHD, i.e., the physical resistivity is zero. However, due to the finite grid size in our simulations, there is a non-zero numerical resistivity. This numerical resistivity is approximately $\eta \simeq \Delta v_{\rm A} \Delta l$, where $\Delta v_{\rm A}$ is the variation in the Alfv\'{e}n speed across a grid cell, and $\Delta l$ is the size of the grid cell \citep[e.g.,][]{Bodenheimer07}. At the ISM-ICM interface where the reconnection occurs, we have $\Delta v_{\rm A} \simeq 50 - 200$~km~s$^{-1}$, and $\Delta l = 1.266$ kpc.  The magnetic Reynolds number is $R_{\rm m} = UL / \eta$ ($U$ is the fluid speed and $L$ is the physical scale of the flow); here we have $R_{\rm m} \gg 1$, therefore the overall flow effectively follows ideal MHD with flux freezing.

The effects of numerical reconnection in our simulations are therefore relatively modest, and probably do not affect the overall results significantly. Reconnection due to the classical resistivity of the plasma would be much slower than the numerical reconnection.  However, astronomical observations suggest that reconnection often occurs at much higher rates, with the inflow speeds for reconnecting magnetic field lines being roughly the Alfv\'{e}n speed \citep{Petschek64}, or even much faster due to turbulence \citep{Kowal09}.  Thus, it is unclear whether our simulations overestimate or underestimate the actual reconnection rates.

Another significant difference between the disjoint and continuous field simulations is that the ISM does not expand and is not heated along the magnetic field at early times ($t \simeq 80 - 400$ Myr) -- here, in the $x$ direction. However, at late times, the narrow ISM tail becomes unstable and turbulent, leading to magnetic reconnection and efficient heat flow in both shielded and disjoint cases. Similarly, when $\mathbf{B} \perp \mathbf{v_{\rm ICM}}$ in the continuous field and disjoint field simulations (Figures~\ref{fig:bycond} and ~\ref{fig:bycondshield}), a significant difference between the two runs is the absence of a prominent L-shaped feature in the shield-cond-perp simulation at $t \simeq 240 - 400$ Myr. As in shield-cond-perp, there is a small amount of cool gas that flows through a narrow region due to numerical reconnection. At $t \simeq 800$ Myr in shield-cond-perp, the stripped double winged tail is prominent and has not expanded and evaporated as in the corresponding continuous case; this region is shielded from the ICM and dissipates at a significantly slower rate than in cont-cond-perp.

Further, we characterize the efficiency and effectiveness of anisotropic thermal conduction when the magnetic fields in the ISM and ICM are completely disjoint. Due to numerical reconnection between the spatially disjoint ISM and ICM magnetic fields, there is a small amount of cool gas that flows out when anisotropic conduction is effective, particularly at early times when the ISM has not been ram pressure stripped ($t \lesssim 400$ Myr). The directions along which heat flows are determined by the initial field configuration; when $\mathbf{B} \parallel \mathbf{v_{\rm ICM}}$ cool gas flows along the $x$ axis, and when $\mathbf{B} \perp \mathbf{v_{\rm ICM}}$ cool gas leaks out along the $y$ axis. At these early times, the leaking of cool gas due to reconnection is the only significant consequence of turning on anisotropic thermal conduction in the shield- simulation series. At late times ($t \gtrsim 1000$ Myr), when most gas has already been stripped, the ISM tail becomes unstable to shear instabilities and is turbulent in shield-cond-par, resulting in a tangled magnetic field distribution. Heat flows and diffuses along tangled field lines. The stripped tail in shield-cond-perp is not turbulent at $t \gtrsim 1000$ Myr, resulting in a lower level of heat diffusion within the tail. This diffusion of heat due to reconnection results in a smoother density and temperature distribution of the stripped ISM compared to the shield-nocond-par and shield-nocond-perp simulations; there are no other significant differences in the evolution of the ISM when an initial disjoint magnetic field prevents any heat flow due to anisotropic thermal conduction. 
 
\section{Discussion}
\label{sec:discussion}
\subsection{Anisotropic thermal conduction and the survival of galaxies and coronae}

\begin{figure*}[!htbp]
  \begin{center}    
    {\includegraphics[width=\textwidth]{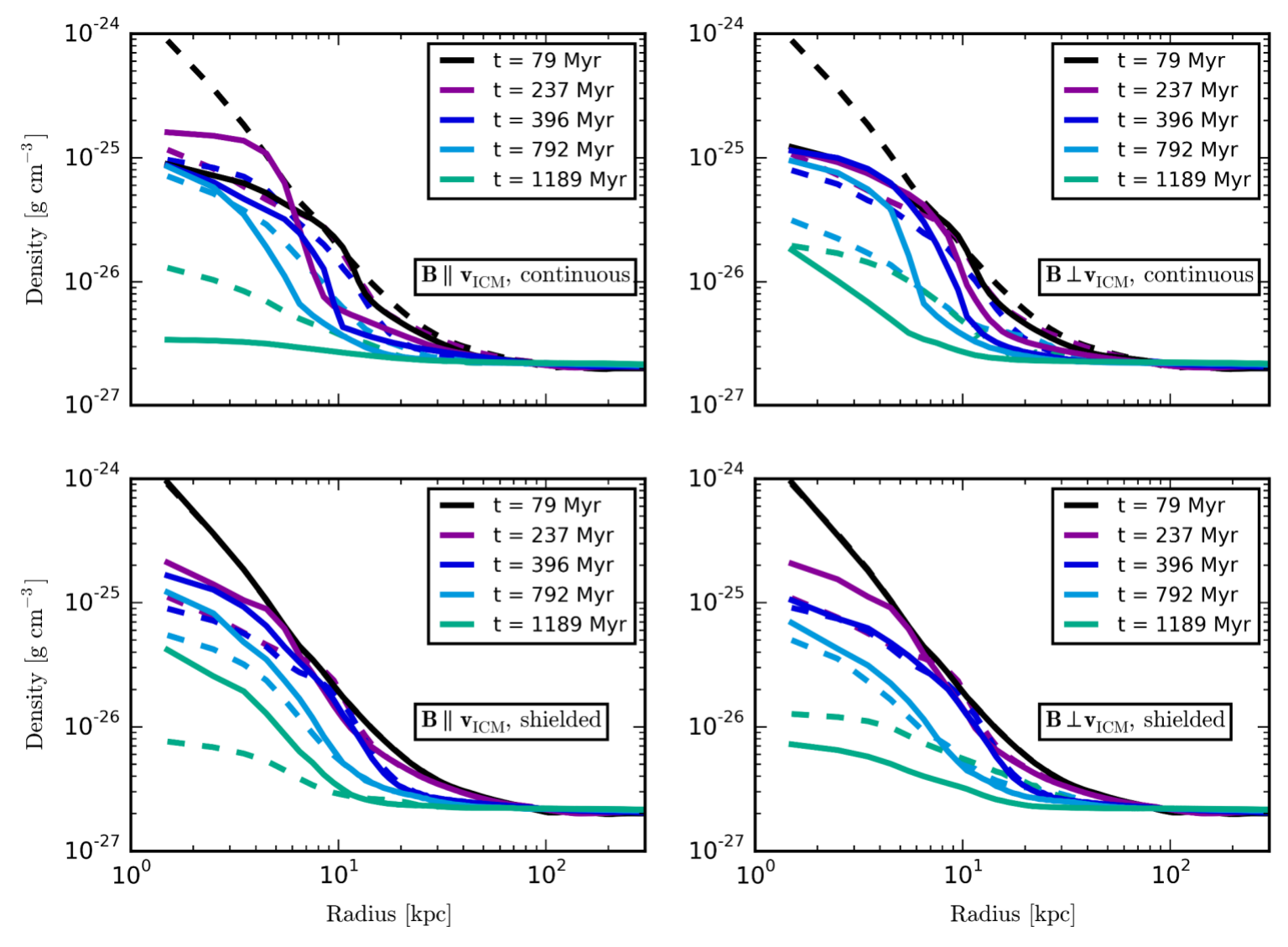}}
     \caption{Evolution of radial gas density profiles. Solid lines correspond to simulations that include anisotropic thermal conduction, and dashed lines to simulations without conduction. 
     The four panels give the results for the four initial magnetic field configurations we consider here.
     \label{fig:densprof_all}}
  \end{center}  
\end{figure*}

\begin{figure*}[!htbp]
  \begin{center}
    {\includegraphics[width=\textwidth]{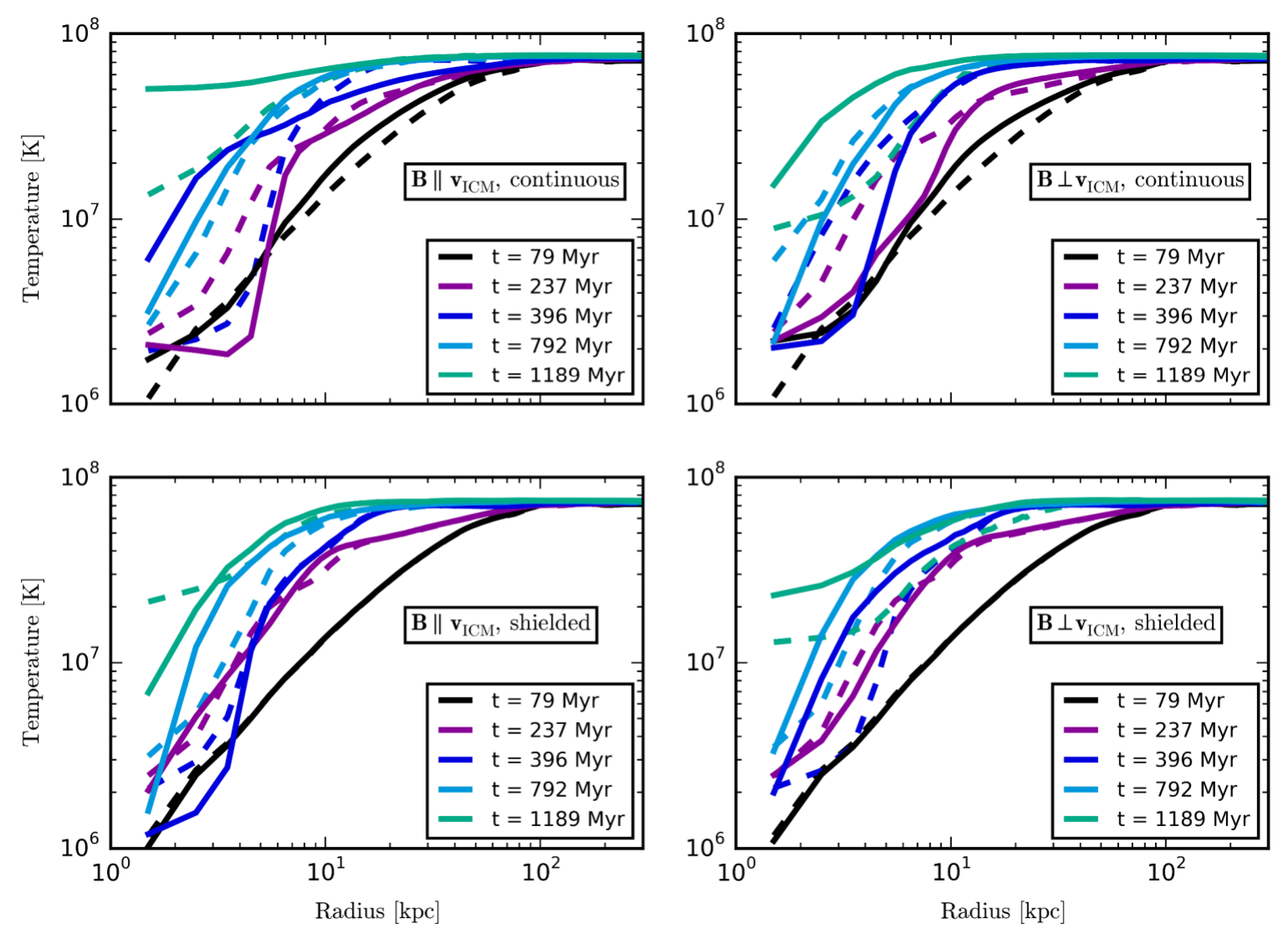}}
     \caption{Evolution of radial temperature profiles. The notation is the same as Fig.~\ref{fig:densprof_all}.
      \label{fig:tempprof_all}}
  \end{center}  
\end{figure*}

In this paper, we simulate the combined effects of ram pressure stripping, a magnetized ISM and ICM, and anisotropic thermal conduction on the evolution of galaxies' hot ISM gas in cluster environments. Magnetic fields alone do not have a significant quantitative effect on ram pressure stripping in cluster environments; under usual ICM conditions, $\beta \simeq 100$. Magnetic pressure is therefore much lower than thermal pressure and ram pressure. Numerical simulations of ram pressure stripping with a magnetized ICM and/or ISM also show that magnetic fields do not significantly affect the mass loss rate of cold disk gas \citep{Ruszkowski14,Tonnesen14} and hot coronal/halo gas \citep{Shin14,Vijayaraghavan17a}. 

We show in Paper I that isotropic thermal conduction alone, in the absence of magnetic fields, results in the rapid evaporation of galaxies hot coronae. The thermal evaporation time is more than an order of magnitude shorter than the ram pressure stripping time. In addition, isotropic thermal conduction qualitatively affects the evolution of the ISM: Kelvin-Helmholtz instabilities are suppressed, and galaxies' diffuse outer coronae evaporate well before they can form ram pressure stripped tails. The presence of long-lived X-ray emitting galactic coronae as well as stripped X-ray tails in cluster galaxies require isotropic thermal conduction to be drastically suppressed from the idealized, saturated, isotropic scenario. ICM magnetic fields force thermal conduction to be anisotropic.

Therefore, although magnetic fields do not quantitatively affect the survival of coronae in the presence of ram pressure stripping, they play an important role in shielding coronae from thermal evaporation. We quantitatively compare the evolution of coronae in the presence and absence of thermal conduction in Figures~\ref{fig:densprof_all} and ~\ref{fig:tempprof_all}. The four panels in each figure show the azimuthally averaged evolution of density (Figure~\ref{fig:densprof_all}) and temperature (Figure~\ref{fig:tempprof_all}) for the four magnetic field configurations in our simulations (cont-par, cont-perp, shield-par, and shield-perp); for each magnetic field configuration, the solid lines show profiles in the presence of anisotropic thermal conduction, and dashed lines the corresponding cases without conduction. We see that in all magnetic field configurations, anisotropic conduction slightly accelerates gas loss, but not at a rate nearly comparable to isotropic conduction.
Density profiles do not differ for a given timestep and field configuration by more than 10--20\% outside the galaxy's core ($r \gtrsim 10 - 20$ kpc). There are differences between the continuous field case and the shielded field case, for both $\mathbf{B} \parallel \mathbf{v_{\rm ICM}}$  and $\mathbf{B} \perp \mathbf{v_{\rm ICM}}$, at early times: when the magnetic field is disjoint, evaporation and the flow of heat is delayed, and density and temperature profiles are identical at $t = 80$ Myr; this is not the case for the continuous field.  

\subsection{Survival of galaxy tails}

\begin{figure*}[!htbp]
  \begin{center}
    \subfigure[cont-cond-perp]
    {\includegraphics[width=0.45\textwidth]{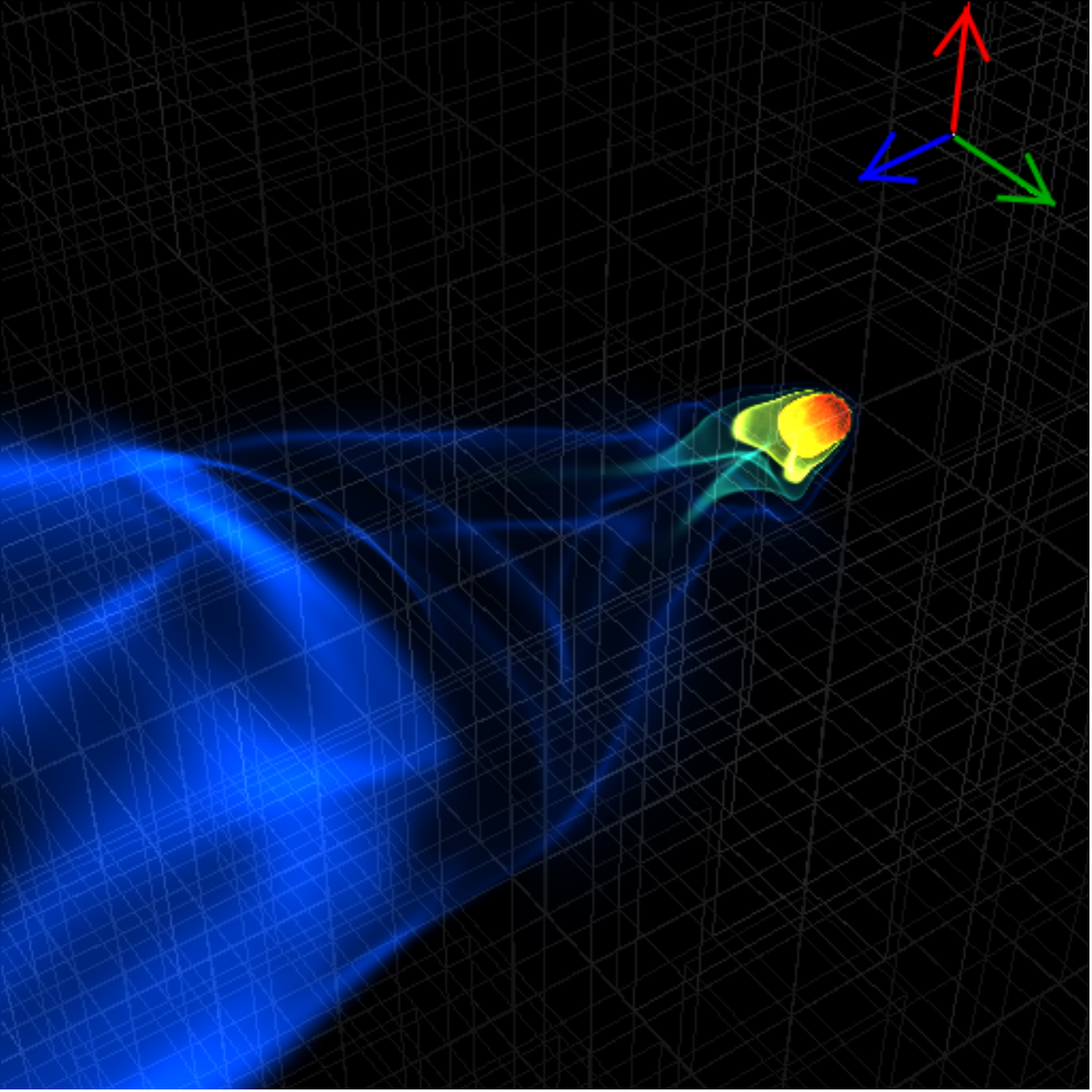}}
    \subfigure[shield-cond-perp]
    {\includegraphics[width=0.45\textwidth]{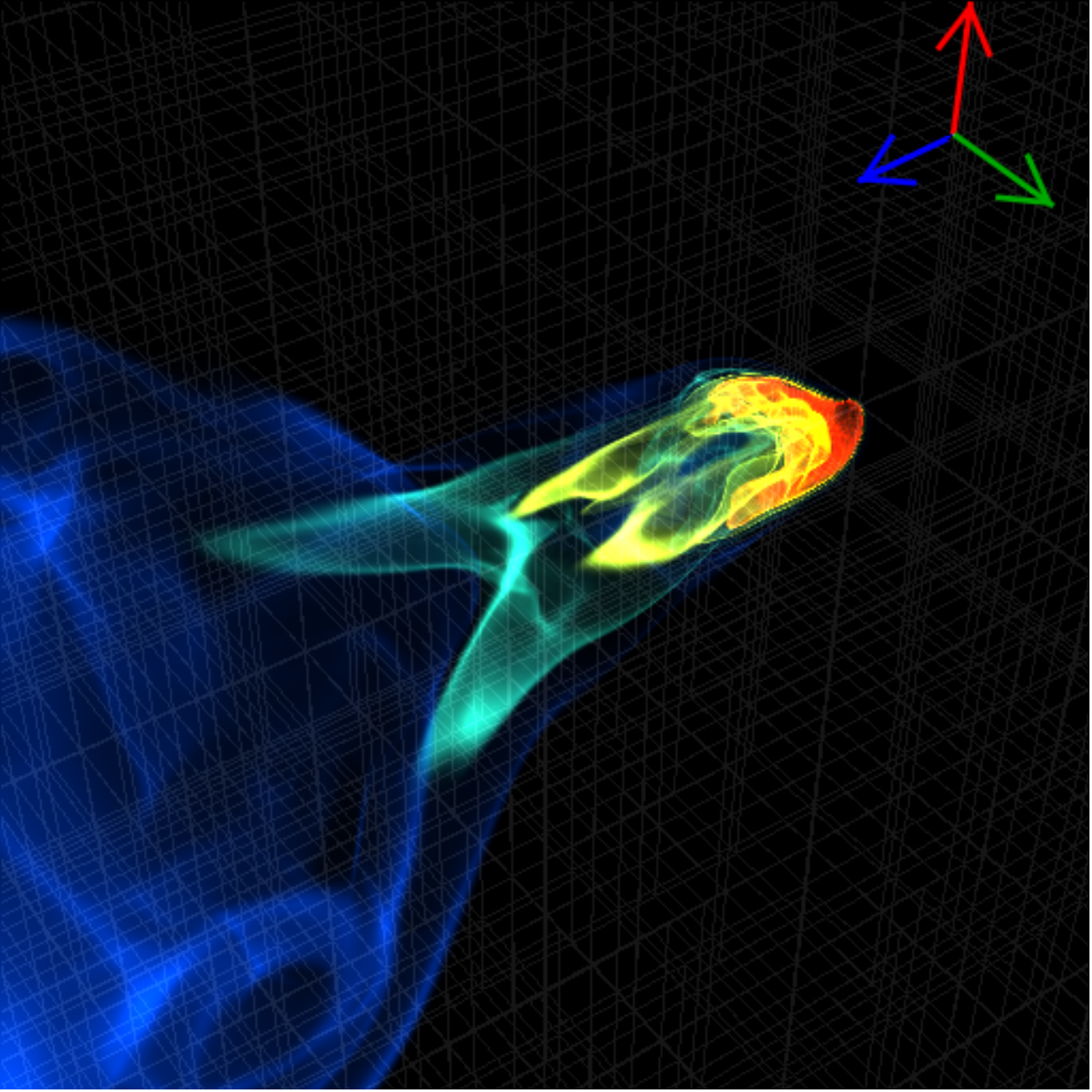}}
     \caption{Three dimensional volume renderings of density at $t = 680$ Myr in the  $\mathbf{B} \perp \mathbf{v_{\rm ICM}}$ simulation with anisotropic thermal conduction, with a connected ICM + ISM magnetic field (left) and   disjoint / shielded ISM-ICM field (right). The upper right hand corner shows the directions of the three axes, where the blue line is the direction of the $x$-axis, the red line is the direction of the $y$-axis, and the green line is the direction of the $z$-axis. Grid lines are overdrawn at the lowest resolution level.
 The center of the dark matter distribution of the galaxy is at the grid vertex associated with the highest gas density in the left panel. Animations for these figures are available.         
      \label{fig:3d_doubletail}}
  \end{center}  
\end{figure*}

When thermal conduction is anisotropic rather than isotropic, it is slowed to the extent that stripped galaxies form tails of gas that do not immediately evaporate. Magnetic fields shield these stripped galaxy tails from thermal evaporation into the hot ICM gas.  The morphology of these stripped tails, and the effectiveness of local magnetic fields in suppressing conduction, depend on the initial field configuration. When $\mathbf{B} \parallel \mathbf{v_{\rm ICM}}$, the tail at late times is narrow and turbulent; when $\mathbf{B} \perp \mathbf{v_{\rm ICM}}$, the tail is
shielded, forming a cone-like structure that does not become turbulent. At intermediate times, ($t \lesssim 800$ Myr), the orientation of the magnetic field (i.e., parallel or perpendicular) affects the ICM wake behind the galaxy.
In the parallel case, the wake is strongly magnetized with a narrow region of increased magnetic field strength.
The ICM magnetic field that is advected around the galaxy converges behind the galaxy, and is swept along by the ICM. In the perpendicular case, the field lines form an
L-shaped structure; they are parallel to the ICM wind in the region where the field lines are advected around the galaxy, and remain perpendicular in other regions. Continuous vs. shielded field configurations also result in tails of different shapes: when the field is shielded,
the tail forms a distinct double-winged structure
(Figure~\ref{fig:3d_doubletail}).
When the magnetic field is continuous and connects the ISM and ICM, heat can flow in a straightforward fashion from the ISM and the ICM, and conduction is more efficient. The presence of an ICM wind flowing around the galaxy bends, enhances, and even results in divergent magnetic fields; the magnetic field is no longer fully continuous and the efficiency of anisotropic conduction is reduced. When the initial magnetic field configurations are disjoint, we see that the flow of wind and advection of field lines,
along with numerical reconnection of field lines,
provides pathways to heat flow between the ICM and ISM. 

We note that a limitation of our simulations is that while we explore a range of magnetic field orientations and morphologies, we do not vary the strength of the magnetic field. The initial ICM $\beta = 160$; for a weaker magnetic field, the lower Alfv\'{e}n speed will result in higher susceptibility to Kelvin-Helmholtz shear instabilities  at the ICM-ISM interface, particularly along the tail. This is because the suppression of these instabilities requires that the relative fluid velocity must be less than the RMS value of the Alfv\'{e}n speeds in the ISM and ICM \citep{Chandrasekhar61} -- and the weaker the magnetic field, the lower the Alfv\'{e}n speed. In this scenario, stripped tails can dissipate more rapidly due to more pathways for heat transport from the ICM. Conversely, a stronger initial magnetic field will result in stronger suppression of shear instabilities, particularly in the $\mathbf{B} \parallel \mathbf{v_{\rm ICM}}$ cases, leading to slower heat transport and longer lived tails. 

\citet{Shin14}, in simulations of ram pressure stripping of elliptical galaxies' hot gas in a magnetized medium, also find similar tail characteristics. They consider an ICM magnetic field that is parallel or perpendicular to the ICM wind, and in all their simulations, the magnetic field is always initially continuous from the ICM to the ISM. They find that in the parallel case, the stripped galaxy tail is elongated in the direction of the wind, while in the perpendicular case the tail is wider along the direction of the field. They also find that when  $\mathbf{B} \perp \mathbf{v_{\rm ICM}}$, the magnetic field is initially amplified along the galaxy's leading edge. \citet{Tonnesen14} simulated ram pressure stripping of a disk galaxy with a magnetized ISM and unmagnetized ICM, and \citet{Ruszkowski14} simulated ram pressure stripping of a disk galaxy with an unmagnetized ISM and magnetized ICM. \citet{Tonnesen14} consider toroidal and dipole-like magnetic fields in the disk.
They do not find significant compression and field enhancement in the disk. They also find that magnetic fields suppress mixing between the ICM and ISM, and that stripped galaxy tails magnetize the ICM. \citet{Ruszkowski14} find that magnetic fields result in long filamentary tails, with magnetic fields aligned with the filaments. They also find that these filamentary tails can occasionally appear as bifurcated structures. 

Under the realistic ICM conditions in our simulations, we find that the timescales for the survival of stripped tails in the presence of anisotropic conduction are similar to equivalent cases without conduction, i.e., when only ram pressure stripping is effective. The tail survival time is much shorter when conduction is isotropic (Paper I); stripped ISM tails are not allowed to form at all as the outer diffuse ISM corona rapidly evaporates.

As described in the introduction, there exist many observed examples of long-lived X-ray emitting stripped tails. \citet{Forman79} observed extended X-ray emission from the stripped tail of M86, an early-type galaxy in the Virgo cluster.  Later \textit{Chandra} observations by \citet{Randall08} showed that this stripped tail is likely almost $400$ kpc long, with possibly a double-tailed structure. \citet{Machacek06} observed an X-ray emitting tail from M89, another early-type galaxy in Virgo. \citet{Irwin96} and \citet{Kraft11} observed the ram pressure stripped tail of NGC 4472, an early type galaxy in Virgo.
\citet{Kim08} observe an X-ray tail from NGC 2619, an elliptical galaxy in a group.
\citet{Sun05b,Sun10,Zhang13} also observe stripped X-ray emitting tails from late-type galaxies. 
Some of the observed tails in both early-type and late-type galaxies appear to be double \citep[e.g.,][]{Randall08,Sun10}, as is true of tails in our simulations with $\mathbf{B} \perp \mathbf{v_{\rm ICM}}$. Some of these double tails are associated with disk galaxies, where other factors may play a role.
In addition to the stripped tails of galaxies, tails have also been observed from infalling groups of galaxies in massive clusters. \citet{Eckert14} observe and study an $800$  kpc long stripped tail from in an infalling group in the Abell 2142 cluster; they explicitly conclude that for this tail to survive evaporation, conduction must be suppressed by tangled magnetic fields.

A simplifying assumption in our simulations is that the magnetic field is initially coherent and continuous; we do not consider the effects of ISM and ICM turbulence, or tangled magnetic fields. Turbulence generally should decrease conduction, since it lengthens the pathlength that electrons have to travel.  Of course, tangled magnetic fields that connect hot and cold regions will be more efficient than organized  magnetic fields that isolate those regions, but that isn't a direct result of turbulence.

\subsection{Similarities with the Persistence of Cold Fronts in Subclusters}

As described in the introduction, simulations show that anisotropic thermal conduction is likely responsible for the persistence of cold fronts in cluster-subcluster mergers. These cold fronts are contact discontinuities between cool, dense, subcluster cores, and hotter, less dense intracluster medium gas. In comparison to kpc-scale galactic coronae, cold fronts are larger, more massive regions; however the physical phenomena responsible for the survival of coronae are also at play in the persistence of these cold fronts. \citet{Ettori00} argued that for cold fronts to survive evaporation in the intracluster medium, thermal conduction must be significantly suppressed. Later simulations have confirmed this hypothesis and are useful comparisons to our galaxy-scale simulations. 

\citet{Asai04} use 2D MHD simulations of a subcluster in an ICM wind, with conduction, to show that magnetic fields initially perpendicular to the direction of the ICM as well as disordered magnetic fields stretch across the subcluster and prevent thermal conduction. \citet{Asai05} show that this is true in 3D simulations as well, although the strength of magnetic field amplification is lower over a three dimensional shock front. Additionally, \citet{Asai07} find that ICM magnetic fields are amplified behind subclusters, even when the initial magnetic field is disordered, as the ICM flow converges behind subclusters. Morphologically, the structure of the magnetically amplified region in their uniform perpendicular field simulations (Figure 7 in \citealt{Asai07}) is similar to the galaxies in our cont-nocond-perp and cont-cond-perp simulations. Additionally, the  magnetically amplified region \emph{behind} their subcluster is also similar to the galaxies in the shield-nocond-perp and shield-cond-perp simulations. The plasma $\beta$ in their simulations behind the subclusters is $\beta \sim 10$, comparable to the value of $\beta$ in the amplified region in our simulations, however their initial plasma $\beta \sim 10^3 - 10^4$ is much higher than ours ($\beta = 160$). Another significant difference is that our lower mass galaxies are susceptible to ram pressure stripping and the formation of tails, while the massive subclusters in the \citet{Asai07} simulations do not form such prominent stripped tails of gas. The maximum spatial resolution  in our simulation is $1.27$ kpc, while the resolution in  \citet{Asai07} is $\sim 10$ kpc. At this resolution, our simulations show the presence of thin magnetically amplified sheets, even at late times, and narrow filaments of dense cool gas along magnetic field lines. \citet{Suzuki13} additionally included the effects of anisotropic viscosity in these simulations; among their results they find that the presence of viscosity reduces the strength of the magnetically amplified region behind subclusters. 

Cool cores in clusters can `slosh', or oscillate about the center of the cluster forming spiral patterned cold fronts; this sloshing is usually triggered by an infalling subcluster. Anisotropic thermal conduction likely helps maintain these regions by suppressing the flow of heat from hotter parts of the ICM to the cooler core gas. \citet{ZuHone13} show using MHD simulations that magnetic fields can be amplified and stretched around sloshing cold fronts, preventing the direct flow of heat from the ICM across cold fronts; they also find that magnetic field lines can  form other pathways connecting the hot ICM and cold subclusters that transport heat forming somewhat less sharp discontinuities in density and temperature. This results in some heating of the cool core gas. Similarly we find that while the core of the galaxy and its stripped tails are largely shielded from evaporation, when compared to the case with isotropic thermal conduction, magnetic fields in the stripped tails can, depending on their orientation and morphology, form paths for heat flow both within the stripped ISM and from the ICM to the ISM. \citet{ZuHone15} include the effects of anisotropic viscosity and anisotropic conduction in sloshing cold fronts. They find that while a viscous medium suppresses shear instabilities and therefore forms sharper cold fronts, anisotropic thermal conduction still reduces the sharpness of these cold fronts. In our simulations, shear instabilities do form at late times in the stripped tail, resulting in more efficient heat transport; it is likely that the including viscosity can prevent the formation of such instabilities while heat conduction and ram pressure will still result in the disruption of these tails.

\section{Conclusions}
\label{sec:conclusions}

In Paper I, we presented simulations of hot gas loss in ICM-like conditions due to ram pressure and isotropic, potentially saturated, thermal conduction. We find that in typical ICM conditions, galaxies lose their hot ISM in $\sim$ $10^2$ Myr due to conduction. The diffuse outer corona evaporates before it can form stripped tails, and the cooler denser central coronal gas rapidly expands and evaporates as well due to the inflow of heat from the ICM. Here, we perform simulations of hot gas loss from galaxies in clusters due to anisotropic thermal conduction in a magnetized intracluster medium and interstellar medium. Heat is only allowed to flow along directions parallel to the local magnetic field. The objective of these simulations is to characterize the shielding of galaxies from evaporation by magnetic fields, and quantify the survival of stripped tails. We simulate two extreme magnetic field orientations: parallel and perpendicular to the ICM wind. For each ICM magnetic field orientation, we consider two cases: the magnetic field is continuous between the ICM and the ISM (and galaxy), and the magnetic field in the ICM wraps around, or shields the galaxy while the galaxy's ISM has a toroidal magnetic field. For each of these four orientations, we perform simulations both with and without anisotropic thermal conduction. 

Overall, we find that gas loss rates are at most 10--20\% higher due to the combined effects of anisotropic conduction and ram pressure, when compared to ram pressure stripping alone.  Ram pressure is therefore the dominant gas removing dynamical process here. Magnetic fields drastically restrict the flow of heat in clusters, since electrons cannot easily cross field lines; the gyroradii of ICM electrons are $\sim 10^8$ cm. Under identical ICM conditions and galaxy properties, isotropic conduction results in gas loss occurring $\sim$10 times faster than ram pressure. 

The morphology of stripped tails and the orientation and strength of the magnetic field are sensitive to the initial magnetic field distribution. When the magnetic field is initially parallel to the direction of the ICM wind (or the direction of motion of the galaxy), the ICM field diverges around the galaxy's leading edge and converges behind the galaxy, resulting in a strongly magnetized wake behind the galaxy. The galaxy's stripped tail is long and narrow, eventually becoming turbulent after ram pressure stripping. 

When the initial magnetic field is perpendicular to the ICM wind, the magnetic field at the galaxy's leading edge is compressed, amplified, and is draped around the galaxy's surface. As the ICM wind drags along field lines around the galaxy, the magnetic field is advected with the wind, and is therefore bent at right angles where the advected field lines rejoin the ICM flow. The magnetic field at the galaxy's leading edge is compressed and amplified perpendicular to the direction of the ICM wind. In the presence of anisotropic conduction, heat flows along magnetic field lines. There is almost no transfer of heat from the ICM to the ISM given that the magnetic field is nearly perpendicular to the thermal gradient. Filaments of cool dense gas flow from the galaxy into the ICM in the galaxy's tail and the ICM wake; the shapes of the filaments correspond to the underlying, amplified or advected magnetic field structure.  

When the initial ICM magnetic field wraps around the galaxy and is disjoint from the ISM magnetic field, it shields the ISM from heat transport. There is some flow of heat from the ICM to the ISM at the surface of the galaxy, the stripped tails, and the galaxy's trailing edge. When the stripped tail becomes increasingly more turbulent, numerical magnetic reconnection within the stripped tail and ICM wake create more pathways for heat to flow and the tails to evaporate in the simulations.
Since this is an artifact of the simulations, it is unclear what would occur astrophysically.

We therefore find that magnetic fields that force thermal conduction to be anisotropic effectively suppress conduction and the subsequent rapid evaporation of the hot ISM. This phenomenon can potentially help explain the survival of compact X-ray coronae in cluster galaxies under conditions where where evaporation due to isotropic conduction would be expected to be efficient. The longevity of stripped tails that do not evaporate and dissipate into the ICM is also likely due to shielding by draped and aligned magnetic fields. One caveat is that these simulations consider only magnetic field configurations which are initially orderly on the scale of the galaxy, and not turbulent fields.

\section*{Acknowledgments}
RV was supported by an NSF Astronomy and Astrophysics Postdoctoral Fellowship under award AST-1501374 and partially by the NASA Chandra theory award TM5-16008X. CLS was supported in part by NASA Chandra grants GO5-16131X and GO5-16146X and NASA XMM-Newton grants NNX15AG26G, NNX16AH23G, and NNX17AC69G. The simulations presented here were carried out on the Rivanna cluster at the University of Virginia. FLASH was developed largely by the DOE-supported ASC/Alliances Center for Astrophysical Thermonuclear Flashes at the University of Chicago. The figures in this paper were generated using the \texttt{yt} software package \citep{Turk11}. We are grateful to the anonymous referee for useful comments that helped improve the discussion in this paper. 

\bibliography{ms}

\end{document}